\documentclass{emulateapj-rtx4}
\usepackage{apjfonts}
\usepackage{amssymb}
\usepackage{natbib}
 \usepackage{color}

\usepackage{graphicx,mathrsfs,amsmath}

\shorttitle{Detection of cool galactic wind at $\lowercase{z} \sim 1$}
\shortauthors{Bordoloi et al.}

\begin{document}
\title{The dependence of Galactic outflows on the properties and orientation of \lowercase{z}COSMOS galaxies at $\lowercase{z} \sim 1$\altaffilmark{+}
}
\author{R. Bordoloi\altaffilmark{1,2}
  , S. J. Lilly\altaffilmark{2}
, E. Hardmeier\altaffilmark{2}
   ,T. Contini\altaffilmark{5,6}
  , J. -P. Kneib\altaffilmark{7}
  , O. Le Fevre\altaffilmark{7}
  , V. Mainieri\altaffilmark{8}
  , A. Renzini\altaffilmark{9}
  , M. Scodeggio\altaffilmark{10}
  , G. Zamorani\altaffilmark{3}
    , S. Bardelli\altaffilmark{3}
  , M. Bolzonella\altaffilmark{3}
  , A. Bongiorno\altaffilmark{11}
  , K. Caputi\altaffilmark{13}
    , C.M. Carollo\altaffilmark{2} 
  , O. Cucciati\altaffilmark{12}
  , S. de la Torre\altaffilmark{13}
  , L. de Ravel\altaffilmark{13}
  , B. Garilli\altaffilmark{10,7}
 , A. Iovino\altaffilmark{4}
, P. Kampczyk\altaffilmark{2}
  , K. Kova\v{c}\altaffilmark{2}
  , C. Knobel\altaffilmark{2}
  , F. Lamareille\altaffilmark{5,6}
  , J. -F. Le Borgne\altaffilmark{5,6}
  , V. Le Brun\altaffilmark{7}
  , C. Maier\altaffilmark{18}
  , M. Mignoli\altaffilmark{3}
   , P. Oesch\altaffilmark{17}
   , R. Pello\altaffilmark{5,6}
  , Y. Peng\altaffilmark{2}
  , E. Perez Montero\altaffilmark{5,6,14}
  , V. Presotto\altaffilmark{4}
   , J. Silverman\altaffilmark{15}
  , M. Tanaka\altaffilmark{15}
  , L. Tasca\altaffilmark{7}
  , L. Tresse\altaffilmark{7}
  , D. Vergani\altaffilmark{3}
  , E. Zucca\altaffilmark{3}
   , A. Cappi\altaffilmark{3}
  , A. Cimatti\altaffilmark{16}
  , G. Coppa\altaffilmark{11}
  , P. Franzetti\altaffilmark{10}
  , A. Koekemoer\altaffilmark{1}
   , M. Moresco\altaffilmark{16}
  , P. Nair\altaffilmark{2}
   \&  L. Pozzetti\altaffilmark{3}
 }

\email{bordoloi@stsci.edu}
 
 \altaffiltext{+}{based on observations undertaken at the European Southern Observatory (ESO) Very Large Telescope (VLT) under Large Program 175.A-0839}
 \altaffiltext{1}{Space Telescope Science Institute, Baltimore, MD 21218, USA}
\altaffiltext{2}{Institute for Astronomy, ETH Z\"{u}rich, Wolfgang-Pauli-Strasse 27, 8093, Z\"{u}rich, Switzerland}
\altaffiltext{3}{INAF Osservatorio Astronomico di Bologna, Bologna, Italy}
\altaffiltext{4}{INAF Osservatorio Astronomico di Brera, Milan, Italy}
\altaffiltext{5}{Institut de Recherche en Astrophysique et Plan\'{e}tologie, CNRS, 14, avenue Edouard Belin, F-31400 Toulouse, France }
\altaffiltext{6}{IRAP, Universit\'{e} de Toulouse, UPS-OMP, Toulouse, France}
\altaffiltext{7}{Laboratoire d'Astrophysique de Marseille, CNRS/Aix-Marseille Universit\'{e}, 38 rue Fr\'{e}d\'{e}ric Joliot-Curie, 13388, Marseille cedex 13, France}
\altaffiltext{8}{European Southern Observatory, Garching, Germany}
\altaffiltext{9}{Dipartimento di Astronomia, Universita di Padova, Padova, Italy}
\altaffiltext{10}{INAF - IASF Milano, Milan, Italy}
\altaffiltext{11}{Max Planck Institut f\"{u}r Extraterrestrische Physik, Garching, Germany}
\altaffiltext{12}{INAF-Osservatorio Astronomico di Trieste, Trieste, Italy}
\altaffiltext{13}{SUPA, The University of Edinburgh, Royal Observatory, Blackford Hill, Edinburgh EH9 1BD, UK}
\altaffiltext{14}{ Instituto de Astrof\'{i}sica de Andaluc\'{i}a, CSIC, Apartado de correos 3004, 18080 Granada, Spain}
\altaffiltext{15}{Institute for the Physics and Mathematics of the Universe (IPMU), University of Tokyo, Kashiwanoha 5-1-5, Kashiwa-shi, Chiba
277-8568, Japan}
\altaffiltext{16}{Dipartimento di Astronomia, Universit\`{a} degli Studi di Bologna, Bologna, Italy}
\altaffiltext{17}{Hubble Fellow; UCO/Lick Observatory, Department of Astronomy and Astrophysics, University of California, Santa Cruz, CA 95064}
\altaffiltext{18}{University of Vienna, Department of Astronomy, Tuerkenschanzstrasse 17, 1180 Vienna, Austria}

\begin{abstract}
We present an analysis of cool outflowing gas around galaxies, traced by Mg II absorption lines in the co-added spectra of a sample of 486 zCOSMOS galaxies at $\rm{1 \leq z \leq 1.5}$. These galaxies span a range of stellar masses ($\rm{9.45 \leq \log_{10}[M_{*}/M_{\odot}] \leq 10.7}$) and star formation rates ($\rm{ 0.14 \leq \log_{10}[SFR/M_{\odot}yr^{-1}] \leq 2.35}$). We identify the cool outflowing component in the Mg II absorption and find that the equivalent width of the outflowing component increases with stellar mass. The outflow equivalent width also increases steadily with the increasing star formation rate of the galaxies. At similar stellar masses the blue galaxies exhibit a significantly higher outflow equivalent width as compared to red galaxies. The outflow equivalent width shows strong effect with star formation surface density ($\rm{\Sigma_{SFR}}$) of the sample. For the disk galaxies, the outflow equivalent width is higher for the face-on  systems as compared to the edge-on ones, indicating that for the disk galaxies, the outflowing gas is primarily bipolar in geometry. Galaxies typically exhibit outflow velocities ranging from $\rm{-200\;kms^{-1}}$  $\rm{\sim -300 \;kms^{-1}}$ and on average the face-on galaxies exhibit higher outflow velocity as compared to the edge-on ones. Galaxies with irregular morphologies exhibit outflow equivalent width as well as outflow velocities comparable to face on disk galaxies. These galaxies exhibit minimum mass outflow rates $>$ 5-7 $\rm{M_{\odot} yr^{-1}}$ and a mass loading factor ($\rm{ \eta \; = \;  \dot{M}_{out} /SFR}$) comparable to the star formation rates of the galaxies.

\end{abstract}

\section{Introduction}
In today's accepted paradigm of galaxy evolution, galactic-scale outflow plays an important role in carrying material out of the galaxy and is a crucial mechanism that participates in regulating the gas reservoir in the galaxy. This could also be one of the primary mechanisms of removing gas from the galaxies and subsequently quenching the star formation in them. The frequency of occurrence of outflows in galaxies and their dependence on the host galaxy properties such as stellar mass and star formation rates are important ingredients in modeling the evolution of the gas in galaxies and the evolution of the intergalactic medium itself (\citealt{Veilleux2005} and references therein). Recent observational evidence also suggests that the galactic winds give rise to the strong Mg II absorption line systems found within 40-50 kpc of the host galaxies observed in the spectra of background quasars and galaxies \citep{Bordoloi2011a,Kacprzak2011b,Bouche2011,Bordoloi2012a}.

Physical models describing mechanisms that drive galactic outflows have been in the literature for a few decades. In case of an ``energy driven'' wind scenario, supernovae explosions or AGN feedback heat the surrounding gas and sweep up gas from the ISM which gets entrained in the expanding bubbles. These super-bubbles experience Rayleigh-Taylor (RT) instability as they approach the scale height of the disk and begin to fragment and escape out in to the galactic halo in collimated form. These super-bubbles also sweep up cool gas from the ISM which gets entrained within cavities of the hot energy driven outflowing fluid. Another model of ejecting gas from the ISM is the ``momentum driven'' wind scenario, where momentum, imparted by radiation or cosmic ray pressure also carries material out of the galaxy into the galactic halo. However, the relative importance of these two processes is still debated. We refer the reader to \cite{Veilleux2005} and \cite{Heckman2002} for a comprehensive overview of the different mechanisms of galactic outflows. 

In the local universe, the hot outflowing gas, found in the galaxies exhibiting outflows, is observed with X-ray emission, and the cooler phase of the outflowing gas is detected via optical emission lines (e.g. H$\alpha$) and in optical absorption lines against the stellar continuum. In the present day universe, strong winds driven by supernovae are seen primarily in starbursting dwarfs, starburst galaxies, LIRGS and ULIRGS. Winds are frequently found in galaxies having a minimum SFR surface density of $\rm{\Sigma_{SFR} \sim 0.1M_{\odot} yr^{-1} kpc^{-2}}$ \citep{Heckman2002}. Several studies have detected outflows in the local dwarf starbursts and LIRGs up to $z \sim 0.5$ using the Na I D $\lambda \lambda $5890, 5896 doublet and used this blueshifted absorption line to trace the kinematics and column density of the cool 100-1000 K gas entrained in the outflowing gas \citep{Heckman2000,Martin2005,Rupke2005b}.

At much higher redshifts, using UV transitions such as Si II $\lambda$ 1260 and CIV $\lambda \lambda$ 1548, 1550, outflows with velocities of hundreds of $\rm{kms^{-1}}$ have been observed in the spectra of in Lyman Break Galaxies (LBGs)  at $z \sim 3$ \citep{Shapley2003}.  \cite{Newman2012} reported a correlation between outflow strength and $\rm{\Sigma_{SFR}}$ of star forming galaxies at $z \sim 2$. They found a $\rm{\Sigma_{SFR}}$ threshold of $\rm{1 M_{\odot} yr^{-1} kpc^{-2}}$, above which $\rm{M_{*} > 10^{10} M_{\odot}}$ galaxies might have stronger outflows as compared to the low mass galaxies. This value is quite different than the one reported for local galaxies.

Using the Mg II $\lambda \lambda$ 2796, 2803 doublet transitions, \cite{Wiener2009} and \cite{Rubin2010} have studied the properties of star forming galaxies at $z\sim 1.4$ and at $z \sim 0.94$ respectively. They used co-added spectra of large number of galaxies to probe the average outflow properties of the galaxies at those redshifts, seen in absorption against the stellar continuum of the galaxy. They found that the  absorption strength of the outflowing gas exhibits a rise with stellar mass and star formation rate of the galaxies. Additional studies of ``down the barrel'' spectra of individual galaxies at these redshifts have also yielded evidence of redshifted inflowing absorption in Mg II and Fe II transitions \citep{Rubin2011,Kornei2012,Martin2012}. The covering fraction of galaxies with detected (unmasked by outflows) inflowing gas is small ($\sim 6\%$). 

It might also be possible to observe such inflowing system, even in presence of strong outflows, if the direction of the outflowing gas is preferentially perpendicular to the disk axis of the galaxy. Studying the inclination dependence of galactic outflow will yield clues to the direction of the outflow in the host galaxy. At lower redshifts, \cite{chen_outflow2010} have shown that the face-on galaxies show stronger outflows than the edge-on ones, using 150,000 SDSS spectra. At higher redshfits, \cite{Kornei2012} have shown, by dividing their sample into face-on and edge-on samples, that the face-on galaxies on average exhibit higher outflow velocities. However, such studies at high redshifts are rather uncertain due to the difficulty in estimating the inclination of galaxies with clumpy morphologies. 

In a previous paper, using stacked spectra of z > 1 background galaxies to probe absorption of Mg II out to 200 kpc, around $\rm{0.5\leq z \leq 0.9}$ galaxies from the zCOSMOS redshift survey \citep{Lilly2007}, we studied the variation of Mg II absorption equivalent width as a function of galaxy color, mass, environment and azimuthal dependence \citep{Bordoloi2011a}. We found that the MgII halos around blue galaxies have a higher absorption equivalent width as compared to that around red galaxies. At the same stellar mass and amongst the blue galaxies there is a correlation of equivalent width with stellar mass. The absorption falls sharply beyond an impact parameter of100 kpc. Most importantly, we found that at lower impact parameters ($< 40$ kpc) Mg II absorption is much stronger along the projected rotation axis of the galactic disk indicating that such Mg II absorption is due to cool gas entrained in bipolar outflows.

In this work we look at the stacked ``down the barrel'' spectra of $1.0 \leq z \leq 1.5$ zCOSMOS galaxies and investigate the Mg II absorption found in their co-added spectra. Mg II absorption probes the low-ionization cool $T \sim 10^{4}$ K gas entrained in star formation driven galactic outflows.  We use HST/ACS F814W imaging \citep{Koekemoer2007} to estimate the inclination of the galaxies and use SED fitting methods to measure the rest frame colors, star formation rates and mass of the galaxies. One of the advantages of doing this study in the COSMOS field is the extensive photometric information available for each galaxy, and the high signal to noise HST/ACS imaging, which enables us to do an accurate morphological classification of the galaxies under study. 

This paper is organised as follows. In section 2 we first present the spectroscopic dataset that is  used, and describe the selection criteria and the derivation of the final samples of objects. In section 3 we describe the stacking technique used in this study. In section 4 we describe the methods used to estimate the galactic outflow from stacked Mg II absorption feature. In section 5.1 we show the dependence of outflowing gas traced by Mg II absorption on rest frame color and stellar mass of the host galaxy. In section 5.2 we show how the outflow properties vary with star formation rate of the host galaxy. In section 5.3 we show how the outflow strength depends on the $\Sigma_{SFR}$ of the galaxies. In section 5.4 we study the dependence of outflow properties on the apparent inclination of the host galaxies. In section 6 we estimate the minimum mass outflow rates for these galaxies. In section 7 we summarize our findings.

Throughout this paper, we use a concordance cosmology with $\Omega_{m} = 0.25$, $\Omega_{\Lambda} = 0.75$ and $\rm{H_{0} = 70\; km\; s^{-1}\; Mpc^{-1}}$. Unless stated otherwise all magnitudes are given in the AB system. 

\section{Sample Selection}

\subsection{The zCOSMOS Survey}
The galaxies used in this study are selected from the zCOSMOS survey. The zCOSMOS survey \citep{Lilly2007} is a spectroscopic survey carried out in the COSMOS field \citep{Scoville2007}. This study utilizes the brighter part of the zCOSMOS survey known as zCOSMOS-bright \citep{Lilly2009}.

The zCOSMOS-bright survey was carried out with the VIMOS spectrograph on the ESO UT3 8-m VLT. The final sample used here consists of approximately 20,000 flux limited $I_{AB} \leq 22.5$ galaxies covering the full two square degree COSMOS field (Lilly et al. in prep.). At this flux limit, the observed redshift range is $ 0 < z < 1.5$.  zCOSMOS-bright objects were observed with the MR grism using 1 arcsec slits, yielding a spectral resolution of $R \sim 600$ at 2.5 {\AA} $\rm{pixel}^{-1}$. The spectra cover the wavelength range from 5550 {\AA} to 9650 {\AA}. From repeat measurements the average accuracy of individual redshifts has been demonstrated to be 110 $\rm{kms^{-1}}$.

\begin{figure}[t!]
\centering
    \includegraphics[height=7.5cm,width=8.5cm]{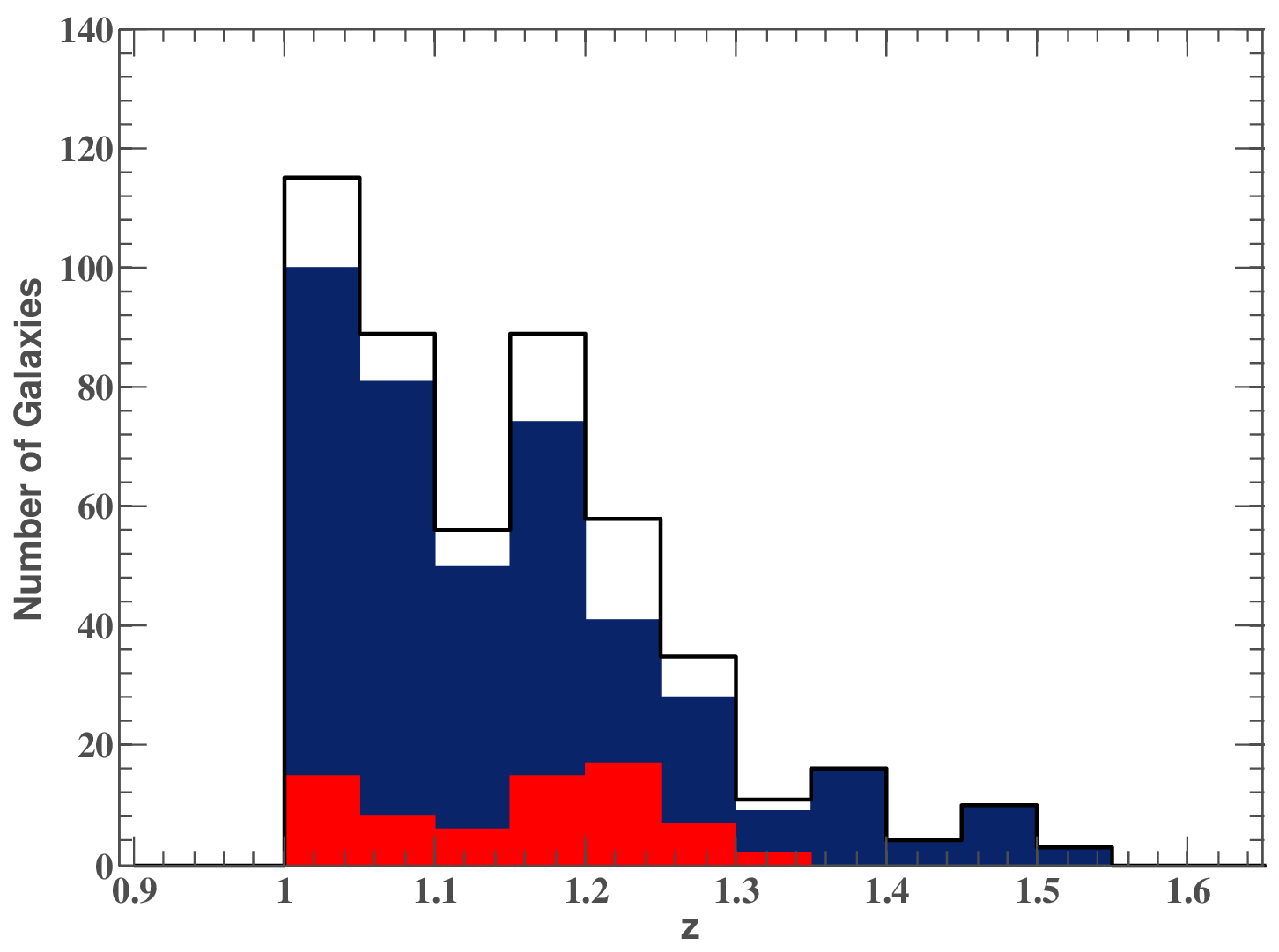}        
    \caption[Redshift distribution of the galaxy sample used in this study]{Redshift distribution of the total galaxy sample (black line) used in the co-add to study the outflowing Mg II gas in bins of $\Delta z = 0.05$. Only spectra with coverage of both Mg II doublet and the O II lines are included. This gives us a  redshift window of $1.0 \leq z \leq 1.5$. The total sample consists of 486 galaxies with $z_{median} \sim 1.1$. The redshift distributions of the red galaxies (red histogram) and the blue galaxies (blue histogram) are also shown.}
\label{fig1}
\end{figure}

\subsection{Selecting the Target Galaxies}

 We first identify the galaxies with secure spectroscopic redshifts, within $1.0 \leq z_{sp} \leq 1.5$. The redshift range is chosen so that the wavelength coverage of the zCOSMOS-bright spectra will cover both the Mg II 2796 {\AA}, 2803 {\AA} absorption doublet as well as the  3727 {\AA} [OII] emission line. This redshift selection yields a sample with 525 galaxies with a median redshift of $z_{median} \sim 1.1$. We utilize the [OII] emission lines to accurately measure the systemic redshifts of the galaxies. About 8\% of the galaxies don't have measurable [OII] emission lines and they are removed from this study. This yields a parent sample of 486 galaxies. Figure \ref{fig1} shows the redshift distribution of these 486 galaxies (black line). The galaxies have redshift Confidence Classes of 4.x, 3.x, 2.5 and 9.5. As shown in \cite{Lilly2009} galaxies with these confidence classes are 99\% reliable in terms of their redshifts. It should be noted that this sample is not mass complete.

\begin{figure}[t!]
\centering
    \includegraphics[height=7.5cm,width=8.5cm]{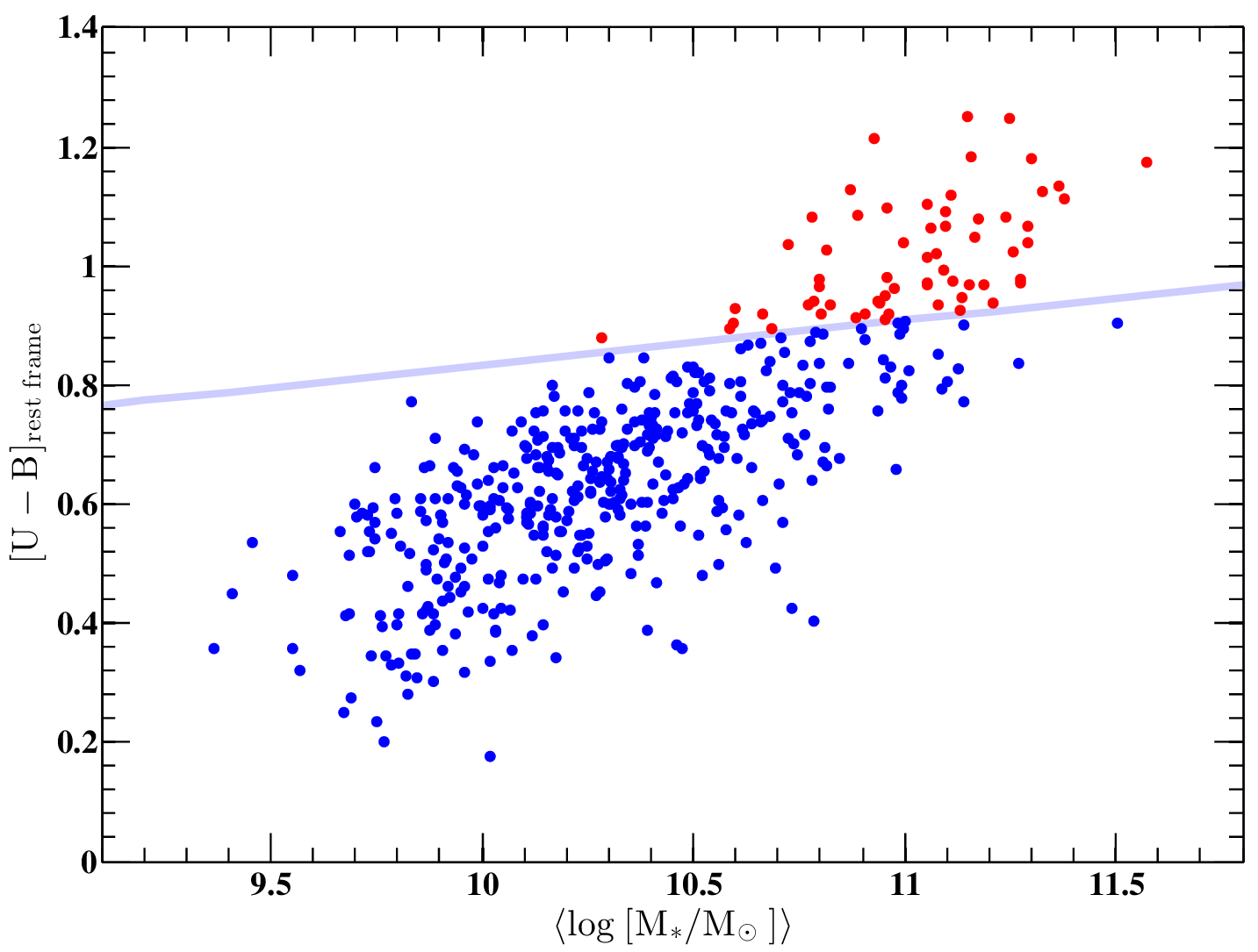}        
    \caption{The target galaxies divided into red and blue galaxies. The red points are for red galaxies and blue points are for blue galaxies respectively. The galaxies are divided into red and blue galaxies using equation \ref{eqn1}. The horizontal line shows the division between red and blue galaxies.}
\label{fig2}
\end{figure}

Each galaxy is associated with a stellar mass and absolute magnitude as described in the previous section. We divide our sample into blue star forming and red passive galaxies with their rest frame (U-B) color, which is a weak function of mass. The line dividing red and blue galaxies is given by 

\begin{equation}
(U-B)_{rest} = 1.05 + 0.075\; \log_{10} \left( \frac{M_{*}}{10^{10} {M_{\odot } }} \right) -0.18z; 
\label{eqn1}
\end{equation}

where $M_{*}$ is the stellar mass of the galaxy in question and $z$ is the redshift of the galaxy. The color-mass division between blue and red galaxies is shown in figure \ref{fig2}. The red points are for red galaxies and blue points are for blue galaxies respectively. After division we are left with 70 red galaxies and 416 blue galaxies respectively. The redshift distribution of the red and blue galaxies are shown in figure \ref{fig1}. The red and blue galaxies essentially have the same N(z). A two sample KS test cannot rule out the null hypothesis that they are drawn from the same parent distribution at 5\% significance.

Stellar masses (and the SFR) for these galaxies have been estimated by spectral energy distribution (SED) fitting, using the Hyperzmass code, a modified version of the photo-z code Hyperz \citep{Bolzonella2010} We refer the reader to \cite{Bolzonella2010} for a detailed technical description of this mass estimation. The absolute magnitudes used in this study were computed as in \cite{Zucca2009}. We use the ZEST morphological classification \citep{Scarlata2007} to estimate the morphologies of the  galaxies. This classification is based on the HST/ACS F814W images in the COSMOS field. We compute the flux averaged star formation rate surface density ($\rm{\Sigma_{SFR} = SFR / 2 \pi R_{1/2}^{2}}$) of the galaxies by dividing their star formation rates with the area enclosed within their half light radii ($R_{1/2}$). To study the dependence of galactic outflow on the apparent inclination of the host galaxies, we select only disk dominated galaxies.We identify disk galaxies classified as ZEST type 2, exclude the bulge dominated systems and divide the sample of galaxies into three bins of inclination angles ($i$), the face-on galaxies ($0^{\circ} \leq i \leq 40^{\circ}$), the intermediately inclined galaxies ($40^{\circ} \leq i \leq 55^{\circ}$) and the edge-on galaxies ($55^{\circ} \leq i \leq 90^{\circ}$) respectively. This yields a total of 218 disk galaxies. Further we probe the outflows in the galaxies classified in ZEST to have irregular morphologies i.e. galaxies with ZEST type 3 which gives a sample of 174 galaxies. The various samples used are shown later in Table 1.

\begin{figure}[ht!]
\centering
    \includegraphics[height=7.5cm,width=8.5cm]{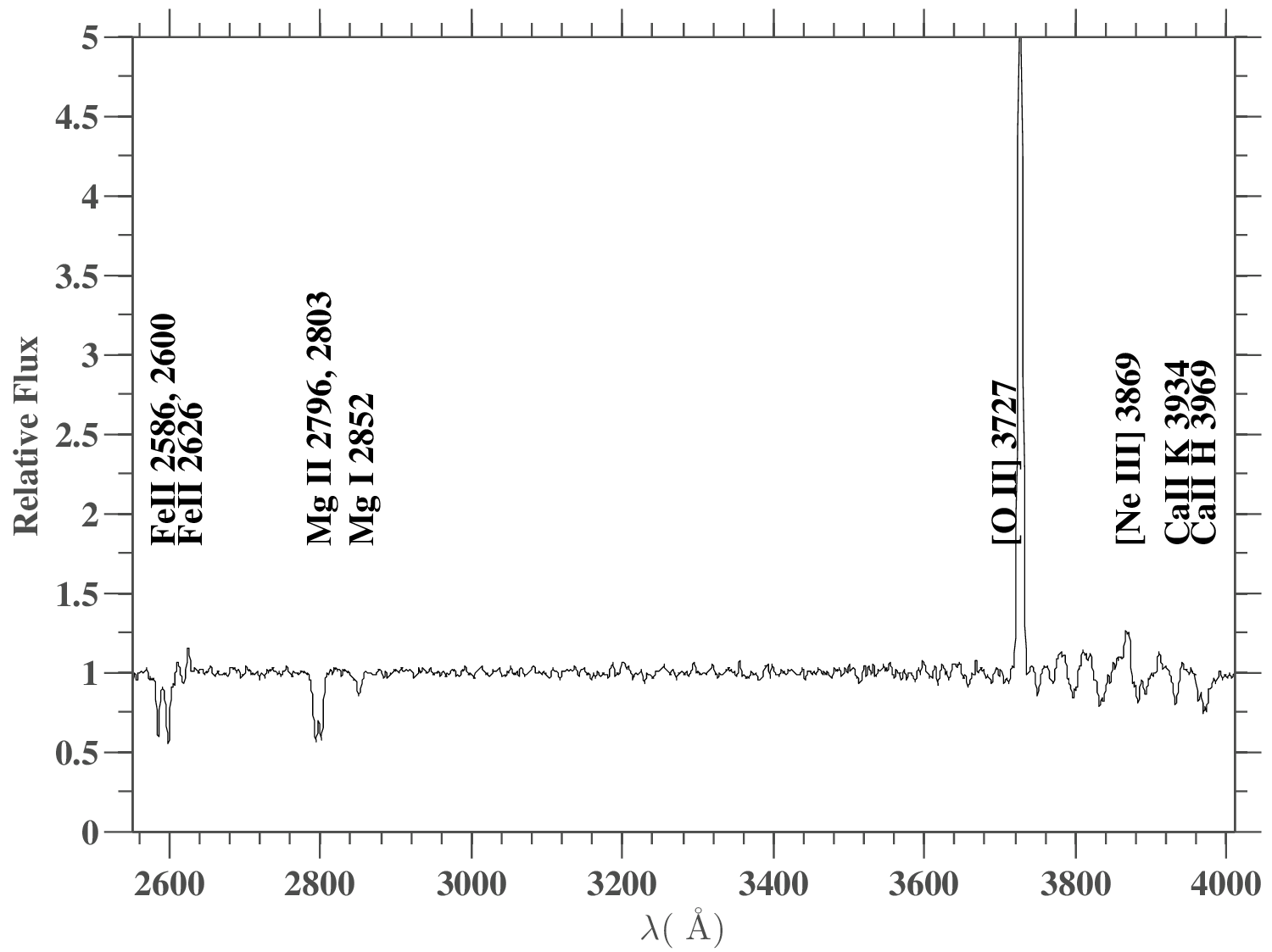}        
    \caption{Co-added spectrum of a sample of 207 $\rm{1.15 \leq z \leq 1.5}$ zCOSMOS galaxies in their rest frame. All the galaxies are continuum normalized to one. The spectrum is smoothed with a 5{\AA} boxcar. }
\label{fig3}
\end{figure}

\begin{figure}[h!]
\centering
    \includegraphics[height=7.5cm,width=8.5cm]{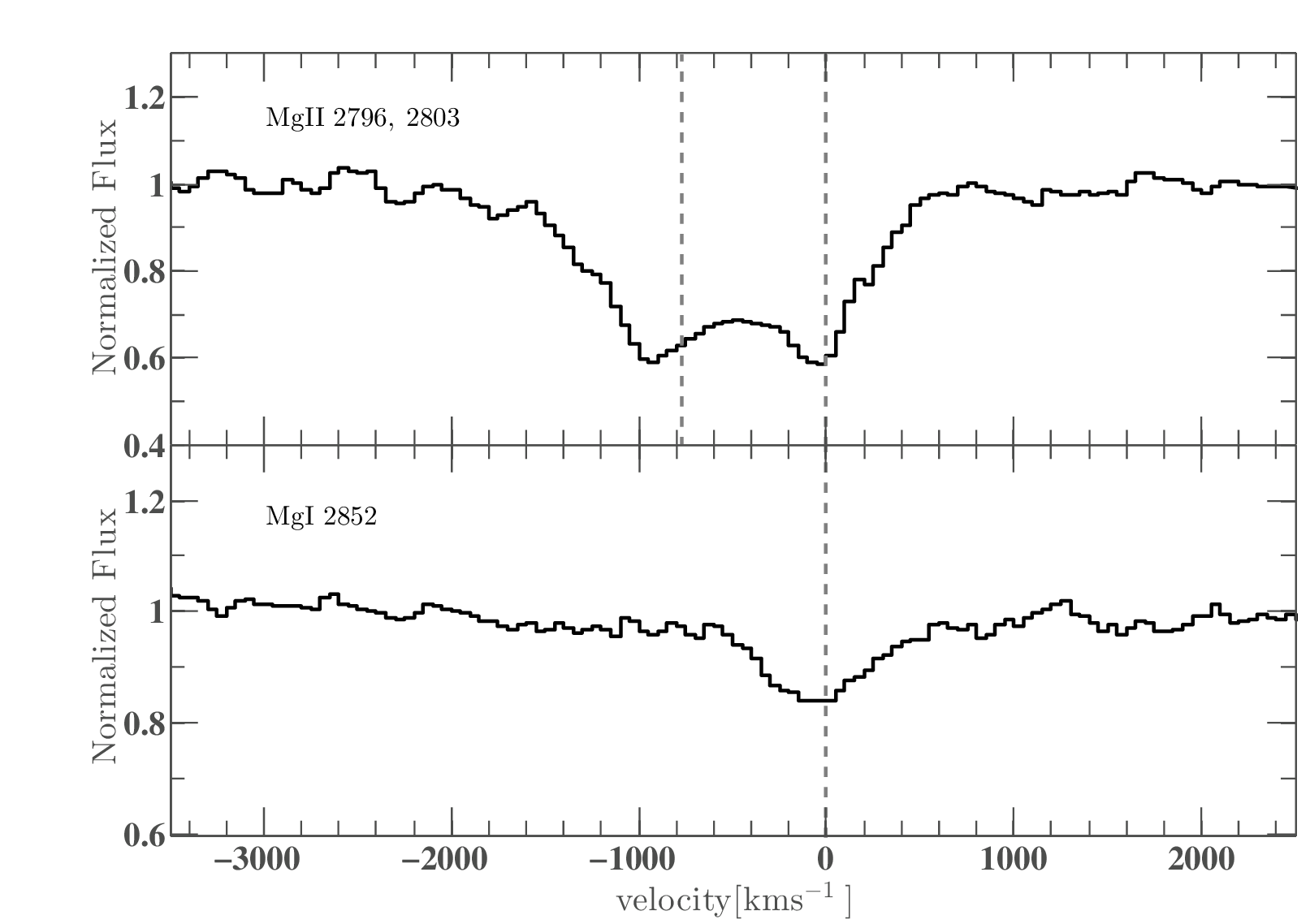}        
    \caption{Mg II 2796, 2803 and Mg I 2852 lines in the co-added spectra of 486 galaxies, relative to the systemic zero velocity as defined by the redshift derived from [O II]. The Mg II and the Mg I lines show blueshifted and asymmetric absorption profiles.}
\label{fig4}
\end{figure}


\section{Co-adding the spectra}

The spectra of all the galaxies in a particular sub-sample, are co-added as follows: the systemic redshift of each individual galaxy is re-determined in a uniform way using the [O II] emission line. Since the spectral resolution of zCOSMOS spectra ($R \sim 600$) is not good enough to resolve the [O II] 3726 {\AA}, 3729 {\AA} doublet, we estimate the systemic redshift of each individual galaxy with detected [O II] emission at the mean wavelength of the [O II] doublet weighted by their line ratios ($j_{3729} / j_{3726} = 0.35$). This gives a line ratio weighted mean wavelength of the [O II] line as 3728.074 {\AA}. Each spectrum is shifted to its rest frame using the new [O II] based systemic redshifts. The rest frame spectra are then resampled on a linear wavelength grid with $\Delta \lambda$ = 0.6 {\AA}, placing all spectra on a common wavelength scale. Each spectrum is then smoothed with a running median box-car filter with a 39 {\AA} smoothing window and a fifth order polynomial is fitted to determine the continuum level, but excluding $\sim$ 40 {\AA} region around the Mg II 2796, 2803 doublet and the [O II] 3727 lines. Each resampled galaxy spectrum is continuum normalized by dividing the galaxy spectrum by this smooth continuum fit described above.

To create the final co-add, a given subset of redshifted, resampled, continuum normalized spectra are co-added by taking the median flux in each wavelength bin. Figure \ref{fig3} shows the co-added spectrum of a subsample of $\rm {\sim 200}$ high z ($\rm{1.15 \leq z \leq 1.5}$) galaxies used in this study, these being chosen here as they also show the FeII 2600, 2587 lines. The co-added spectrum is dominated by strong [O II] 3727 emission. There are prominent Balmer absorption lines and the [Ne III] 3869 emission line is visible. The Ca II H (3969 {\AA}) and Ca II K (3934 {\AA}) lines are also visible. The spectrum blue wards of [O II] 3727 line is relatively featureless except for strong absorption from Mg I at 2852 {\AA} and the Mg II doublet at 2796, 2803 {\AA}. At the bluest end of the spectrum, Fe II 2626 {\AA} emission line is visible and the Fe II $\lambda \lambda$ 2586 {\AA}, 2600 {\AA} doublets are also seen.
In figure \ref{fig4} we extract the regions of the co-added spectrum around the Mg II 2796 {\AA}, 2803 {\AA} doublet and the Mg I 2852 {\AA} line, and plot them on rest frame velocity scales estimated from the [O II] emission line. The vertical dashed lines give the systemic rest frame zero velocity of each individual line. The first visible signature of outflow can be seen here in the sense that, both the Mg II and Mg I lines are asymmetric and blueshifted by a few hundred $\rm{km s^{-1}}$ with respect to the systemic velocity defined by the rest frame redshift of the [O II] line. 

\section{Measuring the outflowing component traced by M\lowercase{g}II absorption}

In this section we describe two methods that are used in the rest of the paper to assess the absorption strength of the blueshifted outflowing gas and the mean velocities. In both methods we estimate the absorption strength of the systemic component by analyzing the red side of the Mg II 2803 {\AA} line. We then estimate the total absorption strength due to the systemic component in the Mg II doublet and ``recover" the absorption strength of the outflowing gas by comparing the systemic component with the observed absorption profile. This ``recovered'' outflowing component is corrected for low spectra resolution effects to yield the final outflow absorption strength. The two methods are described as follows.

\begin{figure}[t!]
\centering
    \includegraphics[height=7.5cm,width=8.5cm]{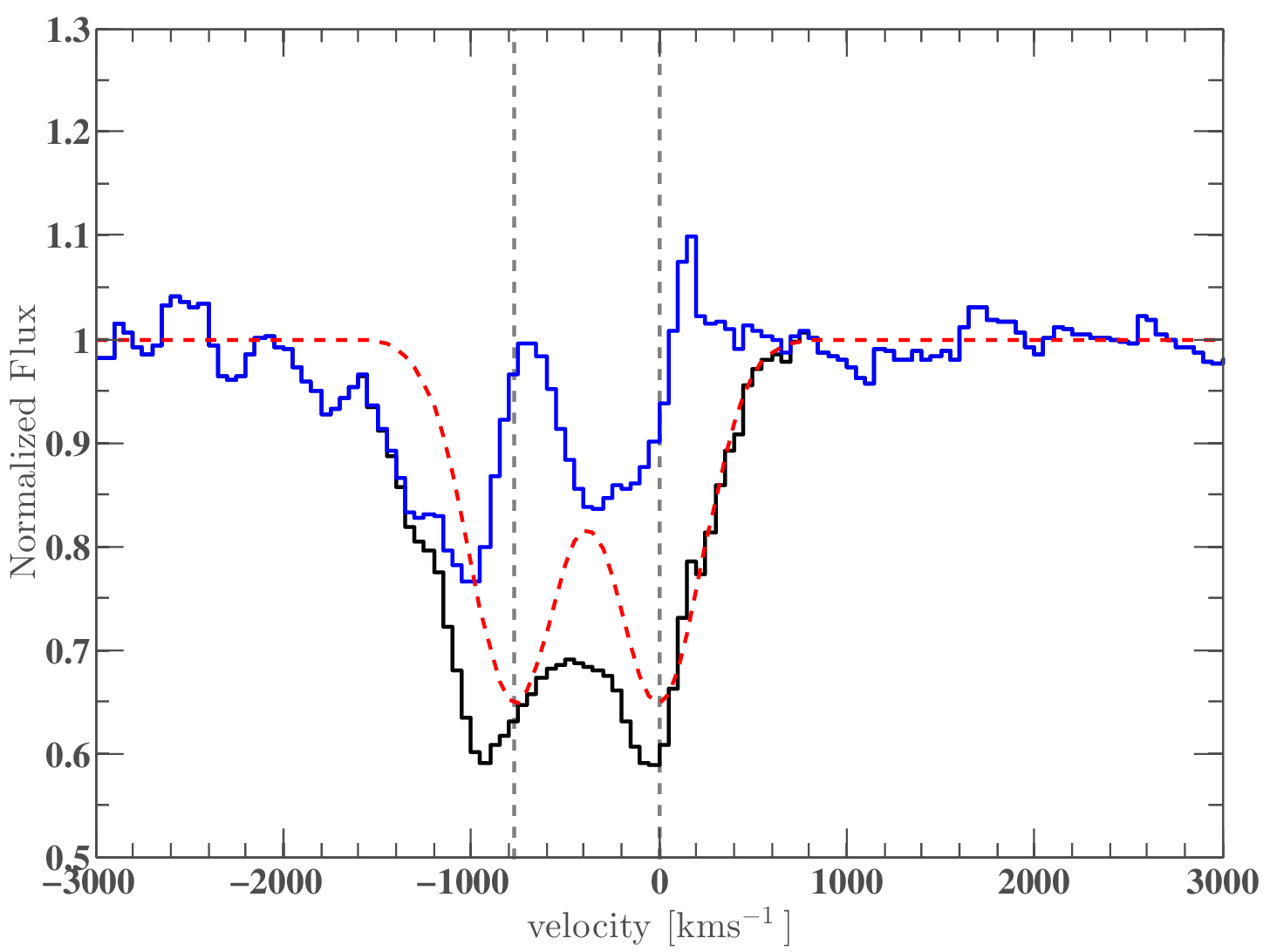}        
    \caption[Example of the decomposition method on the Mg II absorption line]{Sections of the co-add of the full sample around Mg II absorption line. The co-add has been normalized to the level of the continuum surrounding the absorption lines as described in the text and is plotted in thick black. black vertical lines mark the systemic velocity for each transition. The red line overplotted on the co-add shows the Gaussian fits to the red side of the 2803 {\AA} line; the Gaussian model profile for the systemic absorption in the 2803  transition is assumed to describe the systemic absorption in the 2796 {\AA} transition as well. The blue spectrum shows the absorption remaining after the models have been divided out of the co-add.}
\label{fig5}
\end{figure}

\subsection{Decomposition Method}

This method is based on the decomposition method described in \cite{Wiener2009}. We first remove the systemic ISM absorption component from the co-added spectra and then proceed to identify the strength of the outflowing gas in the co-add. The outflowing gas lying between the observer and the light source (the galaxy) will absorb photons at blueshifted negative velocities with respect to the systemic redshift of the galaxy. This outflowing gas could have some velocity dispersion (generally much lower than the thermal broadening of individual clouds) and even complex kinematic structures. However, due to the low resolution of the zCOSMOS spectra, such kinematic complexities can be neglected for the present purposes. The model describing the final observed absorption line is 

\begin{equation}
F_{obs}(\lambda) = C(\lambda) \; F_{em}(\lambda) \; [ 1 - A_{sym}(\lambda) ] \; [1 - A_{flow}(\lambda)]
\label{eqn2}
\end{equation}

\begin{equation}
A_{sym}(\lambda) = A_{2796} \; G(v,\lambda_{2796},\sigma) + A_{2803} \; G(v,\lambda_{2803},\sigma)
\label{eqn3}
\end{equation}

where $F_{obs}(\lambda)$ is the observed flux density from the co-added spectra, $ C(\lambda)$ is the fit to the local continuum of the co-added spectra, $F_{em}(\lambda)$ describes any emission above  the continuum level and for simplicity is set to unity. $[1 - A_{sym}(\lambda)]$  describes the systemic absorption component and $[1 - A_{flow}(\lambda)]$ describes the absorption profile due to outflowing (blueshifted) gas. The symmetric absorption component is taken to be the sum of two Gaussians centered at the rest frame wavelength of each component of the Mg II $\lambda \lambda$ 2796 {\AA}, 2803 {\AA} doublet. 
 
The decomposition of the symmetric and the outflowing component can be a complex problem due to the doublet structure of the absorption profile. We therefore infer the symmetric component by fitting a Gaussian on the red side of the 2803 {\AA} line (within $\rm{0 \;km\; s^{-1}\leq v \leq 1500 \;  km\; s^{-1}}$ ). Since the Mg II doublet is partially blended in our co-add, assuming symmetry we impose the Gaussian fitted to the 2803 line on to the 2796 line as well. We keep the depth of these two Gaussians the same ($A_{2796} =A_{2803} $), as the individual Mg II lines are likely to be saturated. If these absorption lines are not completely  saturated, then it may happen that the 2796 {\AA} line is slightly deeper than the 2803 {\AA} line. In that case we will underestimate the systemic component in that line.

The results of this approach are illustrated in figure \ref{fig5}. The black line is the observed spectrum, the red dashed double Gaussian is the symmetric component $[1-A_{sym}(\lambda)]$, and the blue line is the detected  outflowing component  $[1 - A_{flow}(\lambda)]$.

The total outflow equivalent width is the equivalent width of the outflowing component measured between $\rm{-1500 \;kms^{-1}\;\leq v \leq 0\;kms^{-1}\;}$ with respect to the 2803 {\AA} line and therefore include both the components of the doublet. It should therefore be noted that the quoted outflowing equivalent width throughout the paper is the total equivalent width of the outflowing components from both the 2796 {\AA} and the 2803 {\AA} lines. The mean outflow velocity ($\langle v_{flow} \rangle$) is the mean velocity of the outflowing gas blue ward of the zero velocity of the 2803 {\AA} line up to a velocity chosen adaptively to be the point at which the absorption is absent or minimized and never greater than -768 $\rm{kms^{-1}}$. We also calculate the mean outflow velocity with respect to the 2796 {\AA} line, and these two values are consistent within the uncertainties. Throughout the paper, we quote the outflow velocity with respect to the 2803 {\AA} line. 

The errors on the equivalent width as well as the outflow velocities are estimated using a bootstrap approach. For each set of galaxy spectra to be co-added, a thousand co-added spectra are generated by randomly selecting spectra from that sample. For each of these co-adds, the outflow equivalent width and the mean outflow velocities are estimated. The width of the distribution of the measured equivalent widths and the outflow velocities in these thousand co-adds are taken as representative of the error in the original measurements and should also account for sample variance, continuum uncertainties and the error in fitting the decomposition model. 

This method assumes that the outflowing profile described in equation \ref{eqn2} is only due to absorption from outflowing gas. Any redshifted emission in the 2803 {\AA} line will change the amplitude and the width of the Gaussian fitted to the symmetric component. Emission to the red side of the 2796 {\AA} line will effectively cause us to underestimate the equivalent width of the outflowing gas. It is very hard to account for the emission filling as the strength of Mg II emission lines will depend on the galaxy mass as well as the dust content of the galaxy (see \cite{Kornei2013} and \cite{Martin2012} for a more detailed description of Mg II emission lines).  We are also assuming that there is negligible amount of gas falling into the galaxy. In-falling gas can also contribute to the red wing of the 2803 {\AA} absorption. However, recent observations of individual galaxy spectra have shown that the covering fraction of infalling gas is quite low, approximately $\sim$ 6 \% \citep{Rubin2011,Martin2012}. Moreover, the equivalent widths of the inflowing component is always lower than that of the outflowing component and hence, it can only be observed when there is little or no outflow. Because of these considerations, we shall neglect the inflowing component in the reminder of this study.

It should be noted that this method works optimally for high S/N spectra, as for such cases the symmetric component can be modelled accurately and hence gives a better estimate of the outflowing gas. 

\subsection{Boxcar Method}

To estimate the equivalent width of the outflowing gas, we also use a second method similar to the method  used by \cite{Rubin2010}. This method is more accurate for low S/N co-adds as compared to the decomposition method. We make an assumption that the Mg II doublet is saturated near the systemic velocities, in our spectra, such that the systemic components of 2796 {\AA} and 2803 {\AA} lines have the same depths at $\rm{v = 0 \; kms^{-1}}$. This is a reasonable assumption as the Mg II absorption line becomes saturated at column densities $\rm{\sim\; 10^{14} \; cm^{-2}}$, which correspond to relatively low hydrogen column densities $\rm{> 10^{19}\; cm^{-2}}$ at solar abundance \citep{Rubin2010}. The contribution from the ISM and the stellar atmosphere which likely makes the dominant contribution to the systemic absorption, typically have column densities exceeding these values.  

We measure the outflow equivalent width as follows

\begin{equation}
W_{out} \; =\; 2 \;\left( W_{2796 {\AA},\;blue} \; - W_{2803 {\AA},\;red} \right)
\label{eqn4}
\end{equation}

where
\begin{equation}
W_{2796 {\AA},\;blue} \; =\; \sum\limits_{\lambda=2785{\AA}}^{2796{\AA}} \left(1 - \frac{F_{obs}(\lambda)}{C(\lambda)} \right) \Delta \lambda 
\label{eqn5}
\end{equation}

\begin{equation}
W_{2803 {\AA},\;red} \; =\; \sum\limits_{\lambda=2803{\AA}}^{2806{\AA}} \left(1 - \frac{F_{obs}(\lambda)}{C(\lambda)} \right) \Delta \lambda
\label{eqn6}
\end{equation}

and $F_{obs}(\lambda)$ is the observed co-added spectra and $C(\lambda)$ is the local continuum around the absorption feature. $W_{2803 {\AA},\;red}$ is the amount of absorption due to gas that is not outflowing, this is subtracted from $W_{2796 {\AA},\;blue}$ to avoid overestimating the outflow absorption strength due to inclusion of the systemic component of absorption. This difference is multiplied by 2 to give the total equivalent width of the outflowing component in the doublet $W_{out}$. This method is more reliable for low S/N spectra because as opposed to the decomposition method, this method does not rely on modeling the systemic component directly.

To estimate the mean outflow velocity in this second method, we consider the mean absorption weighted outflow velocity

\begin{equation}
\langle v_{out} \rangle  \; =\; \frac{W_{all}}{W_{out}}   \langle v_{all}\rangle
\label{eqn7}
\end{equation}

since we can assume that the mean velocity of the systemic component ($\langle v_{sym}\rangle$) is zero. $W_{all}$ is the total equivalent width of the Mg II 2796, 2803 doublet and $\langle v_{all}\rangle$ is the mean absorption weighted velocity of the observed absorption line $F_{obs}(\lambda)$ with respect to its systemic velocity. 

The errors of measurement in this method are estimated similar to the decomposition method. Out of the sample of galaxies under study, randomly selected spectra are co-added one thousand times. For each co-add equivalent widths and velocities are measured as described above. The width of the distribution of the measured equivalent widths and the velocities in these thousand co-adds are taken as representative of the error in the original measurements.

\subsection{Correction for Spectral Resolution}
Both methods described above assume that there is no redshifted absorption from the outflowing gas. But we are dealing with finite spectral resolution, which effectively puts some of the absorption from the outflowing gas into the redshifted absorption. This implies that we will overestimate the systemic component and as a result the outflow equivalent widths will be underestimated as the spectral resolution is degraded.

\begin{figure}[t!]
\centering
    \includegraphics[height=7.5cm,width=8.5cm]{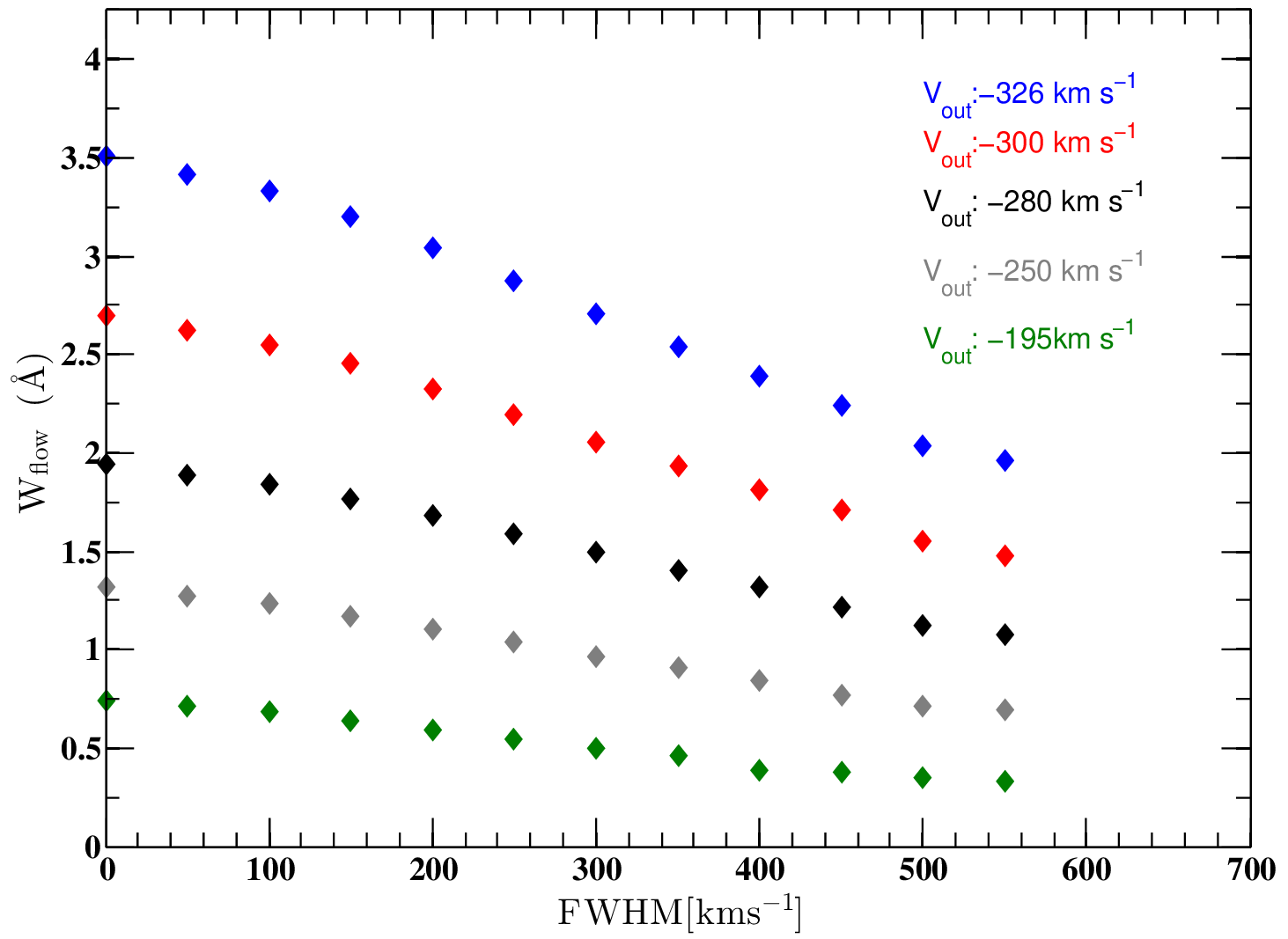} 
        \caption{Variation in estimating the outflow equivalent width as resolution of the spectra is degraded. Five models with different outflow velocities ($\langle V_{out} \rangle$) are considered. The intrinsic outflowing equivalent width ($W_{flow} $ at FWHM=0) is underestimated as the spectra are convolved with the instrumental FWHM.} 
\label{fig6}
\end{figure}

To estimate the typical magnitude of this effect, we set up a simple empirical model, similar to the model described by equation \ref{eqn2} and equation \ref{eqn3}. We set up a systemic absorption component described as $[1 - A_{sym}(\lambda)]$ centered on the Mg II 2796, 2803 absorption doublet. The depth of the two Gaussians are $\sim$ 40\% of the continuum (we assume that the systemic component is saturated and hence the line ratio is 1:1) and $\sigma_{sym} \; =\;100\; \rm{kms^{-1}}$ is chosen. The outflowing component $[1- A_{flow} (\lambda)]$ can be of any arbitrary shape, but we assume that it can also be described by two Gaussian profiles that are blue shifted from the systemic velocity by an arbitrary velocity offset $\rm{\Delta v}$, which is varied. The Gaussians describing the outflowing components are truncated at zero velocity. The depth of the outflowing absorption profile is also varied and the width of the outflowing component varied to be $\rm{\Delta v/2 \leq \sigma_{out} \leq \Delta v}$. We again assume that there is no emission and get an ideal absorption profile $\mathcal{F}(\lambda)$. To account for finite spectral resolution $\mathcal{F}(\lambda)$ is convolved with a Gaussian window function $W$ to give the observed flux $F(\lambda)$ as follows
\begin{equation}
 F(\lambda)\;=\; \left(W * \mathcal{F}\right)(\lambda) \;=\; \int\limits_{-\infty}^{\infty} W(\tau) \mathcal{F}( \lambda - \tau) d\tau.
\label{ch5:eqn8}
\end{equation}

Here $W$ is a Gaussian with a FWHM that is changed to mimic the effect of varying resolution. The equivalent width of the outflow component is then  estimated from $F(\lambda)$ using both the decomposition and the boxcar method. Figure \ref{fig6} shows the observed equivalent width of the outflow component for a set of models with a fixed systemic component and varying outflow component chosen to span an outflow velocity range of $\rm{\rm{-326\; kms^{-1} \leq \langle v_{out} \rangle \leq -195 \; kms^{-1}}}$. For a model with intrinsic outflow equivalent width $\sim 2.7$ {\AA}, the observed outflow equivalent width is $\sim 1.55$ {\AA} for an instrument with a FWHM of 500 $\rm{kms^{-1}}$. This correction needs to be applied to the measurements done with both the decomposition and the boxcar method to estimate the true outflow equivalent width corrected for resolution effects. 

The Mg II absorption doublet from the co-add of all the 486 galaxies was shown in figure \ref{fig5}. Before correcting for resolution, the outflow equivalent width measured from the decomposition method is $W_{flow}\; =\; 1.65 $ {\AA} and the outflow equivalent width measured from the boxcar method is $W_{out}\; =\; 1.76$ {\AA}. Since the resolution of zCOSMOS spectra corresponds to a FWHM of $\sim$ 500 $\rm{kms^{-1}}$, the resolution corrected outflow equivalent widths are $W_{flow}\; =\; 2.55 \pm 0.24$ {\AA} and $W_{out}\; =\; 2.66\pm 0.25$ {\AA} respectively. The outflow velocities inferred by both the methods are $\rm{\langle v_{flow}\rangle \; =\; -343 \pm 49 \; kms^{-1}}$ and $\rm{\langle v_{out} \rangle \; =\; -319 \pm 95 \; kms^{-1}}$ respectively. $\langle v_{out} \rangle $ is estimated using equation \ref{eqn7}. It should be noted that $W_{all}$ and $\langle v_{all} \rangle$ are not affected by the degradation in resolution and hence are robustly estimated from the observed absorption feature without any corrections. As a result, the use of equation \ref{eqn7} to deduce $\langle v_{out} \rangle $ only requires correction of $W_{out}$.

\begin{figure}[t!]
\centering
    \includegraphics[height=7.5cm,width=8.5cm]{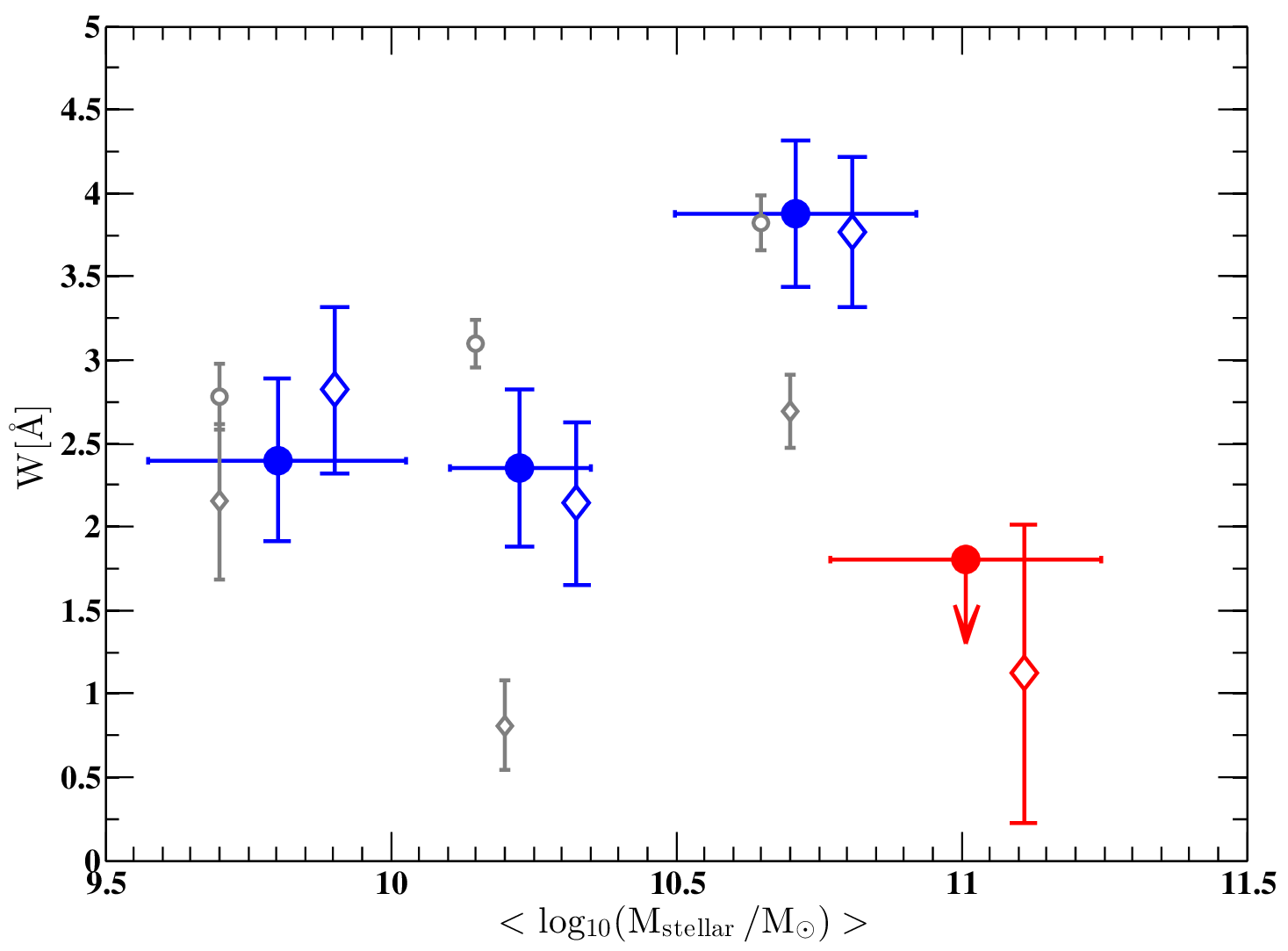} 
      \caption{Variation of the outflow equivalent width as a function of color and mass of the host galaxies. The blue galaxies (blue points)  exhibit a rise in equivalent width  with stellar mass. At similar stellar mass ranges the blue galaxies have significantly higher outflow equivalent width as compared to red galaxies (red point). The measurements done with the decomposition method are (red and blue) filled circles and the measurements done with the boxcar method are open diamonds (red and blue). The gray open circles are measurements from \cite{Wiener2009} and the gray open diamonds are measurements taken from \cite{Rubin2010}. The error bars on the x axis are standard deviation of the masses in each bin. The blue diamonds are offset in mass by 0.1 dex for clarity.}
\label{ch5:fig7}
\end{figure}

\begin{figure}[t!]
\centering
      \includegraphics[height=7.5cm,width=8.5cm]{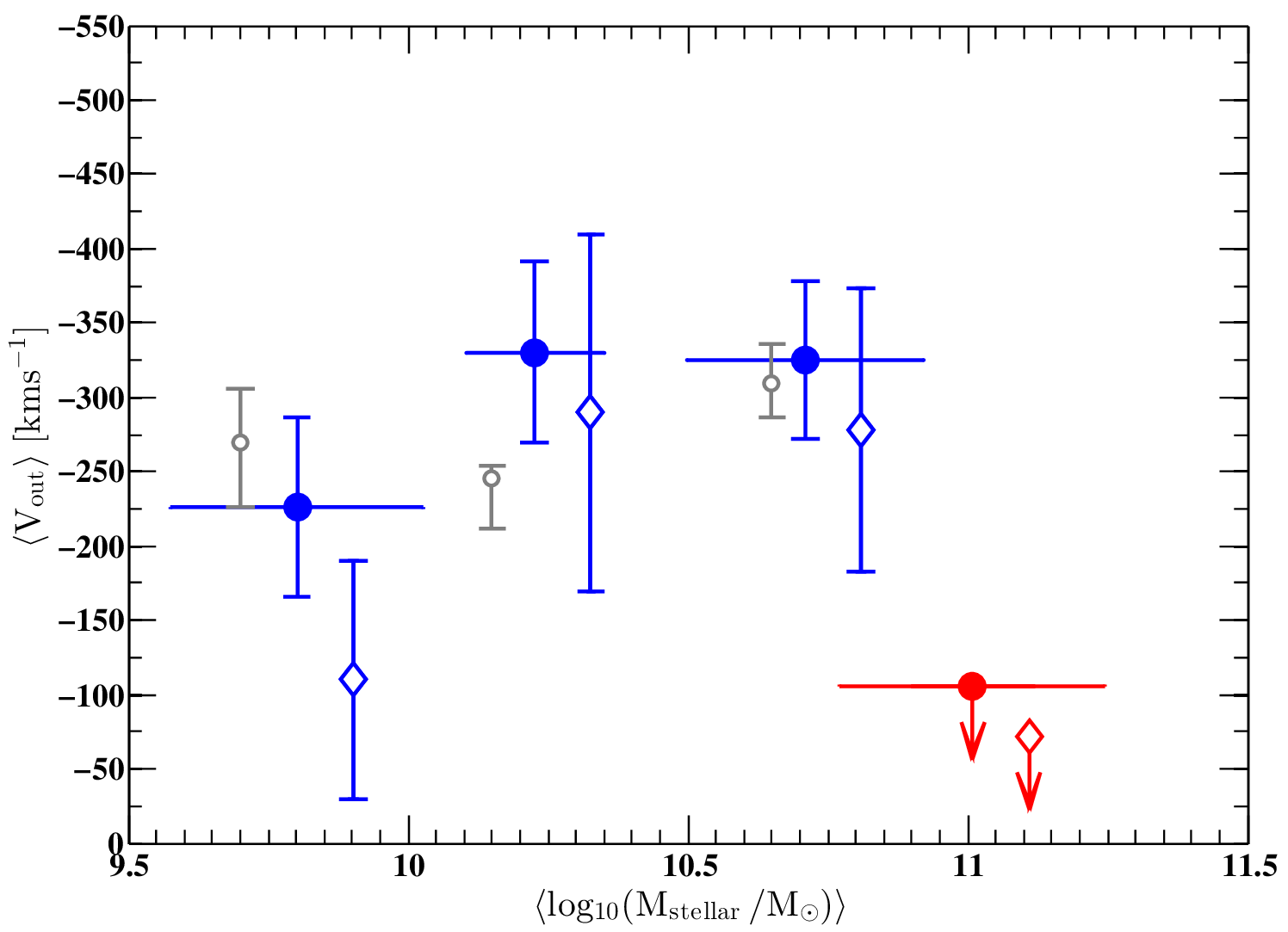}     
    \caption{Variation of outflow velocity as a function of mass of the host galaxies. The measurements done with the decomposition method are filled circles and the measurements done with the boxcar method are open diamonds . The gray open circles are measurements from \cite{Wiener2009}. The error bars on the x axis are standard deviation of the masses in each bin. The blue diamonds are offset in mass by 0.1 dex for clarity.}
\label{ch5:fig8}
\end{figure}

\section{Results}
In the following section we discuss the variation of equivalent widths and velocities of outflowing gas traced by the Mg II absorption with different properties of the host galaxies, examining the dependence on rest frame color, mass, star formation rate, star formation rate surface density ($\Sigma_{SFR}$) and for the disk galaxies, the apparent inclination. The measurements are tabulated in Table \ref{table:sample_table1}.

\subsection{Dependence with Color and Mass}

We first investigate the variation of the outflow equivalent width and outflow velocity with the rest frame color and stellar mass of the host galaxies. The galaxy sample is divided into red and blue galaxies as described before. To study the mass dependence, we divide the blue galaxies into three stellar mass bins. The low mass blue galaxy sample consists of galaxies with $M_{*} \leq 10^{10} M_{\odot}$, the intermediate mass blue galaxy sample consists of galaxies with stellar masses $10^{10} M_{\odot} < M_{*} \leq 10^{10.45} M_{\odot}$ and the high mass blue galaxy sample is defined as galaxies with $M_{*} >10^{10.45} M_{\odot}$

The high mass blue galaxy sample has a mean stellar mass $ \langle M_{*} \rangle = 10^{10.71 \pm .22} M_{\odot}$ which is reasonably close to the mean mass of the red galaxy sample $\langle M_{*} \rangle = 10^{11 \pm .24} M_{\odot}$ to compare the dependence of outflowing properties with color. Figure \ref{ch5:fig7} shows the variation of outflow equivalent width as a function of rest frame color and stellar mass. It can be seen that both the decomposition method (filled circles), and the boxcar method (open diamonds) give similar outflow equivalent widths within the uncertainties. Amongst the blue galaxies, the high mass blue galaxies exhibit higher outflow equivalent width as compared to the low mass blue galaxy samples. At similar stellar mass ranges, the blue galaxies exhibit significantly higher outflow equivalent width as compared to the red galaxies. In the sample of red galaxies, very little outflow absorption is observed. This color and mass dependence in the equivalent width of outflowing component hints at a dependence of the outflowing gas on the star formation rate of the host galaxy. This trend is also similar to the trend observed in \cite{Bordoloi2011a} for impact parameters within 50 kpc of $0.5 \leq z \leq 0.9$ galaxies, seen in the spectra of background galaxies. 

The gray points are data points taken from the literature, the gray open circles are taken from \cite{Wiener2009} and the gray open diamonds are taken from \cite{Rubin2010}. The study done in \cite{Wiener2009} used 1406 spectra of blue star forming galaxies at $z \sim 1.4$ from the DEEP2 galaxy survey. They estimated the outflow equivalent width of the 2796 {\AA} line only and we have multiplied that value by 2 to compare with our estimates of the total outflow equivalent width of the doublet. The study done in \cite{Rubin2010} used a sample of 468 galaxies in the TKRS survey in the GOODS-N field at $0.7 \leq z \leq 1.5$ with a median redshift of 0.9. Both these studies have a higher spectral resolution than the zCOSMOS bright survey. Although these studies were done in different redshift ranges, their general trends are consistent with our finding of a general increase of outflow equivalent width with stellar mass. However, the precise quantitative details of such a correlation are still quite uncertain. Our own analysis highlights the importance of incorporating color in interpreting the trends with mass and/or luminosity.

Figure \ref{ch5:fig8} shows the outflow velocity estimates done with both the decomposition and the boxcar method as a function of stellar mass. For the red galaxies only an upper limit for outflow velocity is measured. The gray points are again taken from \cite{Wiener2009}. We find that within the error bars there is no significant trend in outflow velocity as a function of stellar mass, although the low mass blue sample gives a lower estimate of outflow velocity in the boxcar method as compared to the decomposition method. In general we find a typical range of outflow velocity ranging approximately from -200 to -300 $\rm{kms^{-1}}$ amongst these galaxies. 

\begin{figure}[t!]
\centering
    \includegraphics[height=7.5cm,width=8.5cm]{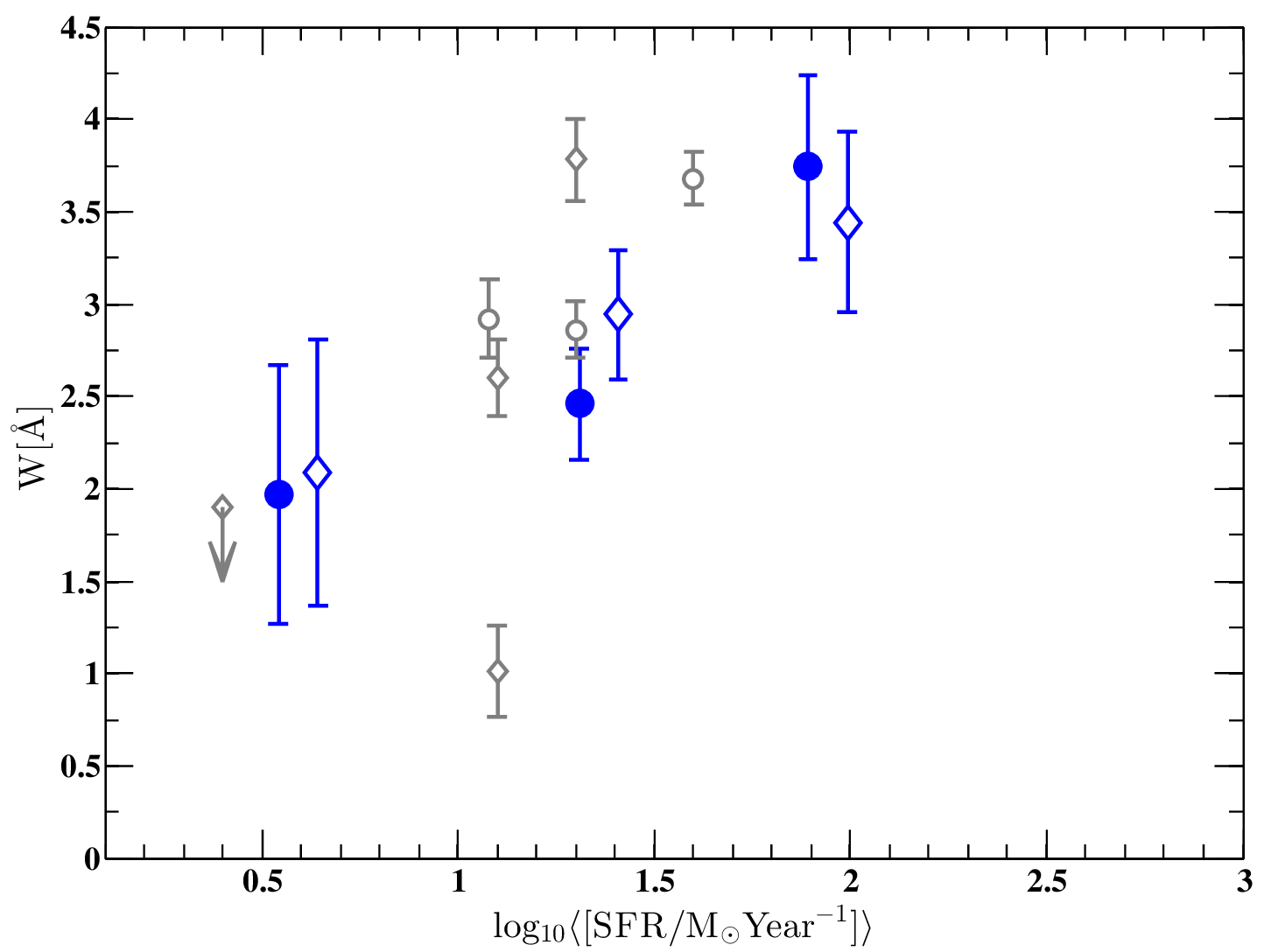}        
    \caption[Variation of outflow equivalent width as a function of star formation rate]{Variation of outflow equivalent width as a function of star formation rate of the host galaxies. The measurements done with the decomposition method are filled circles and the measurements done with the boxcar method are open diamonds. The gray open circles are measurements from \cite{Wiener2009} and the gray open diamonds are measurements from \cite{Rubin2010}. The blue diamonds are offset in SFR for clarity.}
\label{ch5:fig9}
\end{figure}


\subsection{Dependence with SFR}

In this section we investigate the correlation between the star formation rates and the outflow properties of the galaxies. The galaxy sample is sub-divided at the 25th and 75th percentile values of star formation rates. The lowest star forming sample consists of galaxies with $\rm{\log_{10}(SFR / M_{\odot} yr^{-1}) \leq 1.0396}$. The intermediate star forming sample consists of galaxies with $\rm{1.0396 < \log_{10} (SFR / M_{\odot} yr^{-1}) \leq 1.572 }$. The highest star forming sample is made out of galaxies with $\rm{\log_{10} (SFR / M_{\odot} yr^{-1})  > 1.572}$. 

Figure \ref{ch5:fig9} shows the variation of outflow equivalent width as a function of the star formation rate of the galaxies. There is a general trend of steady increase in outflow equivalent width with increasing star formation rates. This trend is seen in both the decomposition method estimates and the boxcar estimates. These trends are consistent with the trends found in literature (gray points) although the other studies were done at different redshift ranges.

\begin{figure}[t!]
\centering
    \includegraphics[height=7.5cm,width=8.5cm]{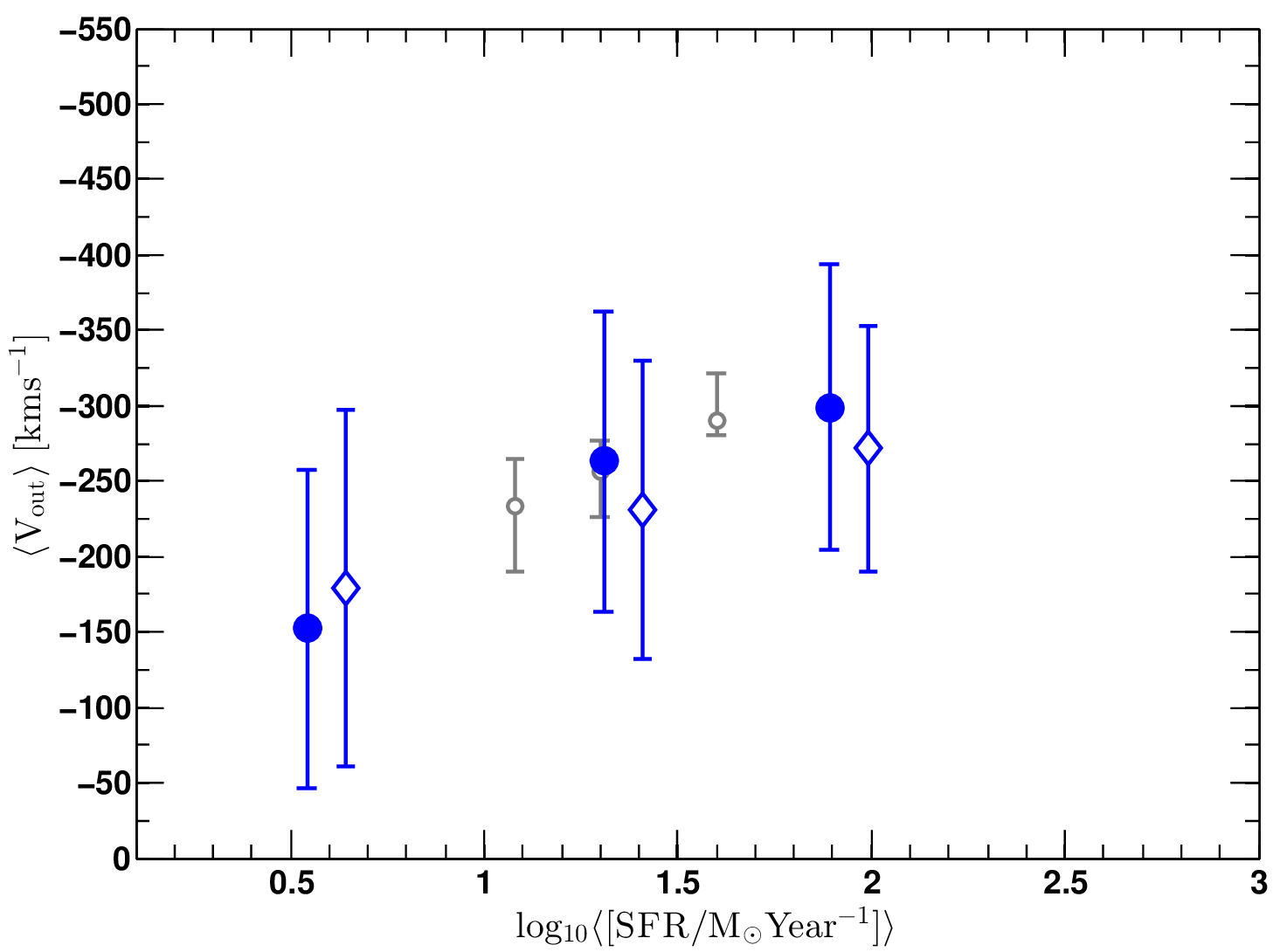}
    \caption{Variation of outflow velocity width as a function of star formation rate of the host galaxies. The measurements done with the decomposition method are filled circles and the measurements done with the boxcar method are open diamonds. The gray open circles are measurements from \cite{Wiener2009} and the gray open diamonds are measurements from \cite{Rubin2010}. The blue circles are offset in SFR for clarity. }
\label{ch5:fig10}
\end{figure}

Figure \ref{ch5:fig10} shows the variation of outflow velocity with varying star formation rates. We find that the mean velocity of the total Mg II absorption ($\langle v_{all} \rangle$) increases with increase of SFR (see Table \ref{table:sample_table1}), the inferred outflow velocities also show a very weak trend of increasing with SFR in the boxcar estimates (open diamonds) whereas they show no such trends in the decomposition estimates (filled circles). However due to large uncertainties in our measurements, both estimates are broadly consistent with the observations of \cite{Wiener2009} (gray circles), who found a weak correlation of increasing outflow velocity with increasing star formation rates. 


\subsection{Dependence on $\Sigma_{SFR}$}


\begin{figure*}[t!]
\centering
    \includegraphics[height=7.5cm,width=8.5cm]{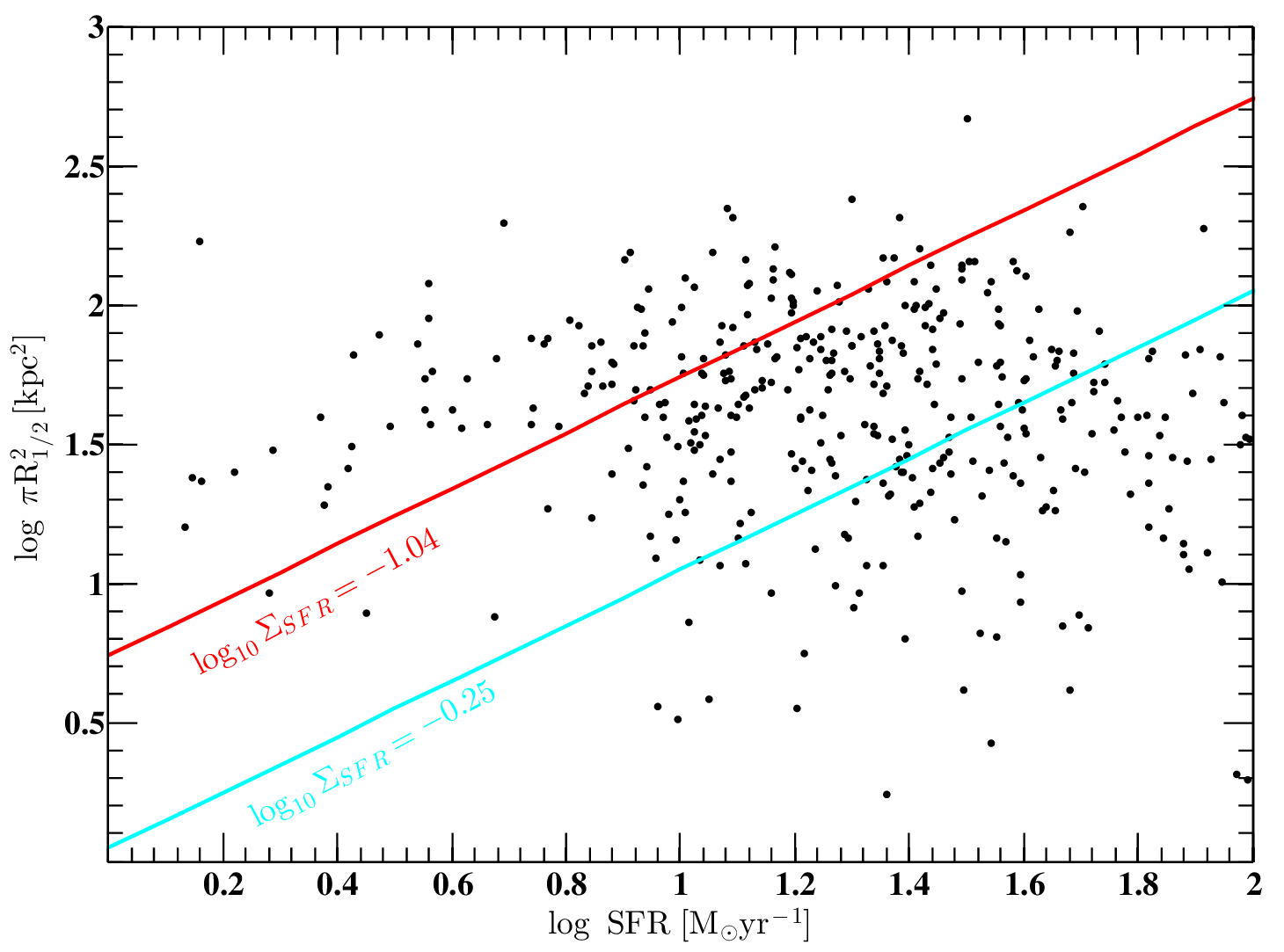}    
        \includegraphics[height=7.5cm,width=8.5cm]{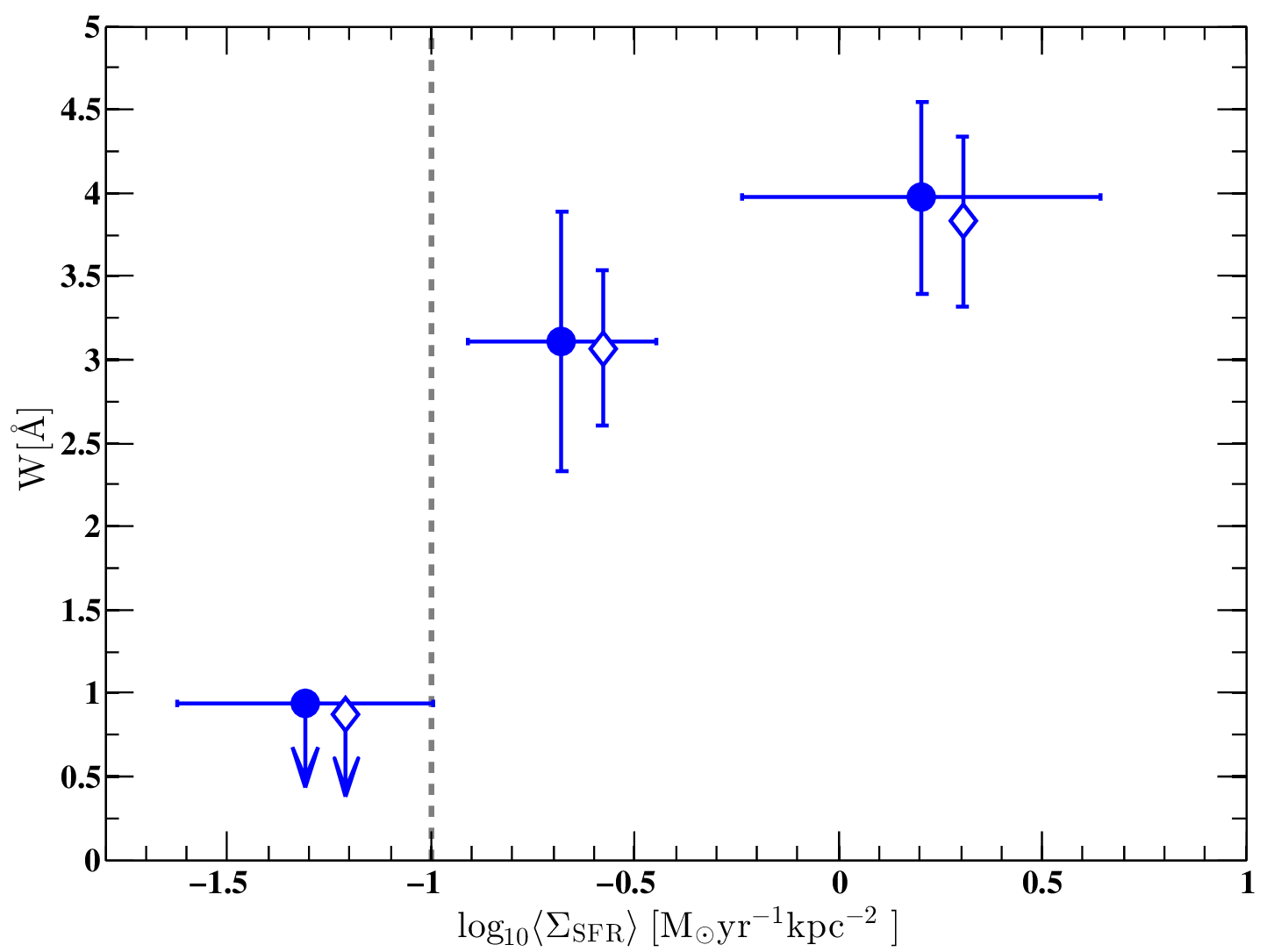}            
    \caption{\textit{Left Panel:-}$\log \pi R^{2}_{1/2}$ vs $\log SFR$ for the target galaxies. Lines show subdivision in $\log \Sigma_{SFR}.$ \textit{Right panel:-} Variation of outflow equivalent width as a function of $\Sigma_{SFR}$. The measurements done with the decomposition method are filled circles and the measurements done with the boxcar method are open diamonds. The blue diamonds are offset in SFR for clarity. The vertical dashed line shows the canonical threshold in star formation surface density for driving winds ($\rm{\Sigma_{SFR} \; = \; 0.1 M_{\odot} yr^{-1} kpc^{-2}}$). Galaxies above this threshold have strong outflow equivalent width but below this threshold, we detect only an upper limit for outflowing equivalent width.  }
\label{fig9}
\end{figure*}

In this section, we test for dependence of outflow strength on the SFR surface density ($\Sigma_{SFR}$). We estimate a flux averaged SFR surface density ($\Sigma_{SFR} = SFR / 2 \pi R_{1/2}^{2}$), for each object assuming that star formation is distributed like the (observed frame) i band light (i.e. half the star forming regions of the host galaxies are within their half light radius ($R_{1/2}$)). Left panel of figure \ref{fig9} shows the distribution of $\log{SFR}$ and $\log{\pi R_{1/2}^{2}}$ for the sample used in this section. The red and cyan lines indicate the 25th and the 75th percentile values of $\log{\Sigma_{SFR}}$ (-1.04, -0.25), which are used to subdivide our sample.

In figure \ref{fig9}, the right panel shows the outflow equivalent width versus $\log{\Sigma_{SFR}}$. The filled circles are measurements done with the decomposition method and the open diamonds are estimated with the boxcar method. The subsample with the lowest $\log{\Sigma_{SFR}}$ exhibit very little outflow equivalent width and as we probe the higher $\log{\Sigma_{SFR}}$ subsamples the measured outflow equivalent width increases. 

It is interesting to discuss these results in the context of the canonical star formation rate surface density ``threshold''  ($\rm{\Sigma_{SFR} \; = \; 0.1 M_{\odot} yr^{-1} kpc^{-2}}$). It has been suggested that it is the minimum $\rm{\Sigma_{SFR} }$ required for driving outflows in local starbursts \citep{Heckman2002}. By chance, this threshold corresponds to the  $\Sigma_{SFR}$ between the first and the second bin in this study, and it is noticeable that the lowest bin has only an upper limit for outflowing equivalent width.

At higher redshifts, \cite{Rubin2010} and \cite{Steidel2010} have reported weaker correlations of outflow absorption strength with $\rm{\Sigma_{SFR}}$. Recently \cite{Kornei2012} reported a 3.1$\sigma$ correlation of outflow velocity with $\rm{\Sigma_{SFR}}$ at these redshifts. At $z \sim 2$, \cite{Newman2012} reported a correlation between outflow strength ( quantified via broad flux fraction of H$\alpha$ line) and $\rm{\Sigma_{SFR}}$ of star forming galaxies. They reported a $\rm{\Sigma_{SFR}}$ threshold of $\rm{1 M_{\odot} yr^{-1} kpc^{-2}}$, above which $\rm{M_{*} > 10^{10} M_{\odot}}$ galaxies might have stronger outflows as compared to the low mass galaxies.

\subsection{Dependence with  apparent inclination angle of disk galaxies}

\begin{figure*}[t!]
\centering
    \includegraphics[height=3.25cm,width=15cm]{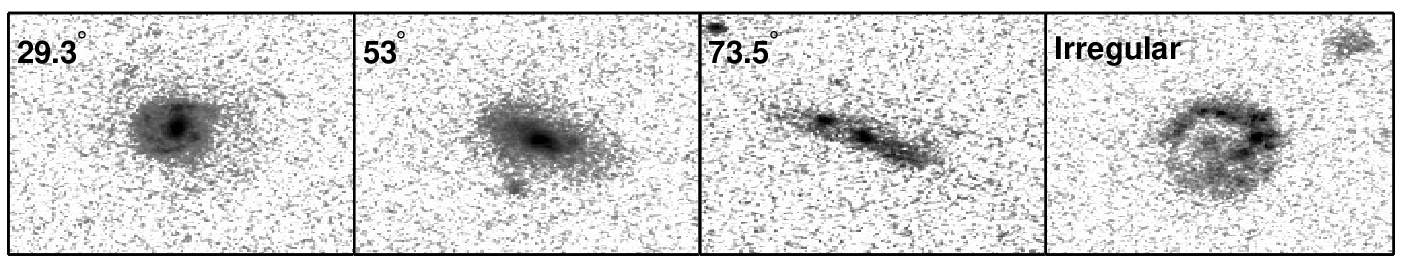}        
    \caption[HST/ACS images of the galaxies with different inclination angles]{HST/ACS images of the galaxies with different inclination angles. The inclination angles are  shown in the top left corner of the stamps. The extreme right galaxy is classified as having an irregular morphology with ZEST type 3. The disk galaxies are chosen not to include bulge dominated systems.}
\label{ch5:fig11}
\end{figure*}

\begin{figure}
\centering
    \includegraphics[height=6.cm,width=8.cm]{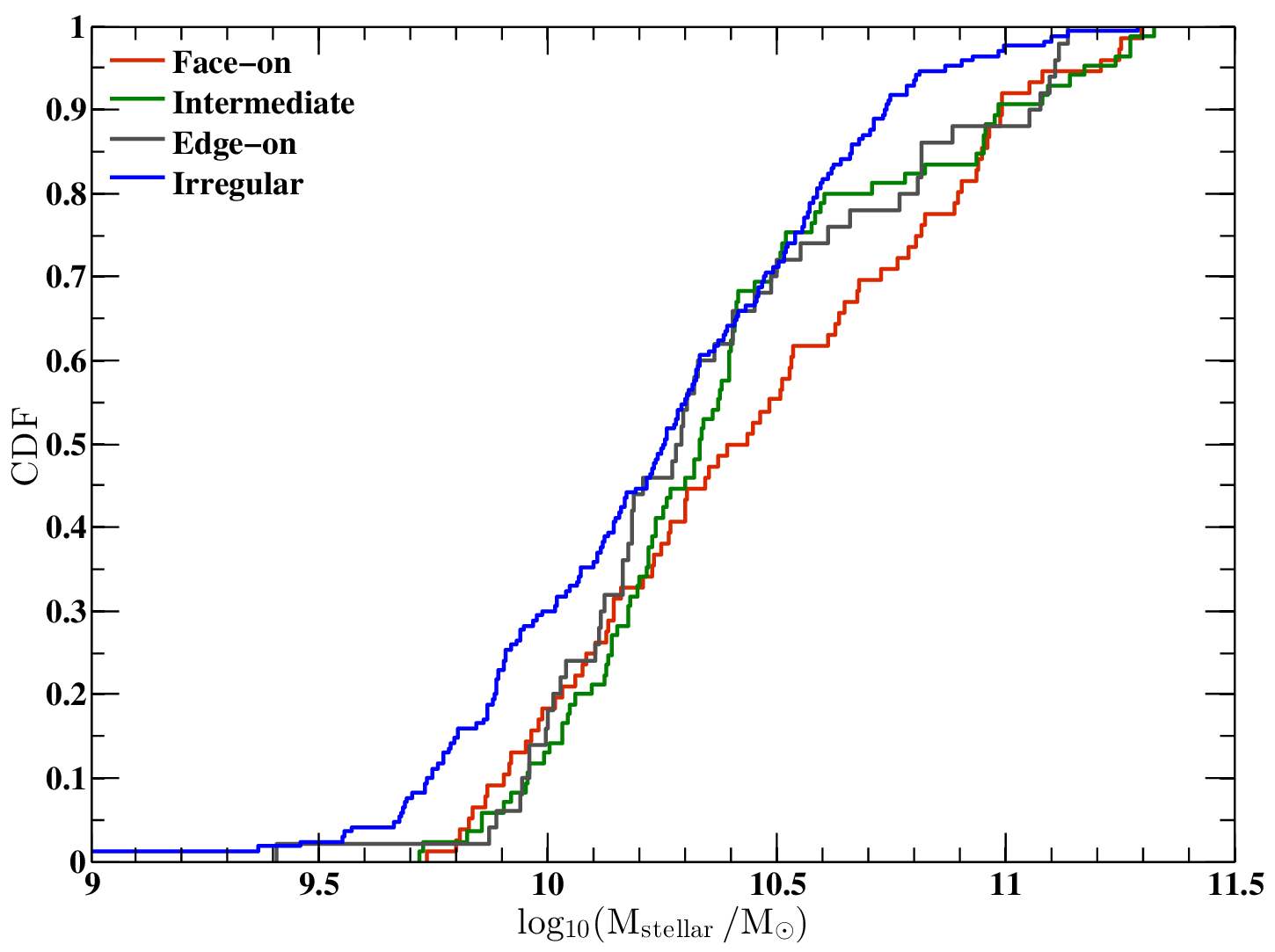}        
    \caption{The cumulative mass distribution of galaxies with different inclinations and irregular morphologies. Two sample KS test performed on the different mass distributions show that the null hypothesis that the disk galaxies are drawn from the same parent mass distribution can not be ruled out at 10\% significance level. The mass distribution of irregular galaxies when compared with that of the disk galaxies shows that the null hypothesis that they are drawn from the same parent mass distribution can be ruled out at 5\% significance level.}
\label{ch5:fig11b}
\end{figure}

In this section we present the variation of Mg II outflow kinematics with the apparent  inclination of the associated disk galaxies.  We select the galaxies  to be disk-dominated, as described before. We divide the disk galaxies into three bins in inclination angles, with $ \rm{0\,^{\circ} \leq i \leq 40\,^{\circ}} $,  $\rm{40\,^{\circ} < i \leq 55\,^{\circ}} $ and $\rm{55\,^{\circ} < i < 90\,^{\circ}} $. The definition of the inclination angle is such that $i = 90\,^{\circ}$ represents an edge-on system and $i = 0\,^{\circ}$ represents a face-on system. Some examples of the typical galaxies selected are shown in figure \ref{ch5:fig11}. We also investigate the mass distribution in each inclination bin and find that they are consistent with having the same mass distribution according to a two sample KS test. The cumulative distributions of the stellar mass of the galaxies in each inclination bin are shown in figure \ref{ch5:fig11b}.

\begin{figure*}[!htb]
\centering
    \includegraphics[height=7.cm,width=8.5cm]{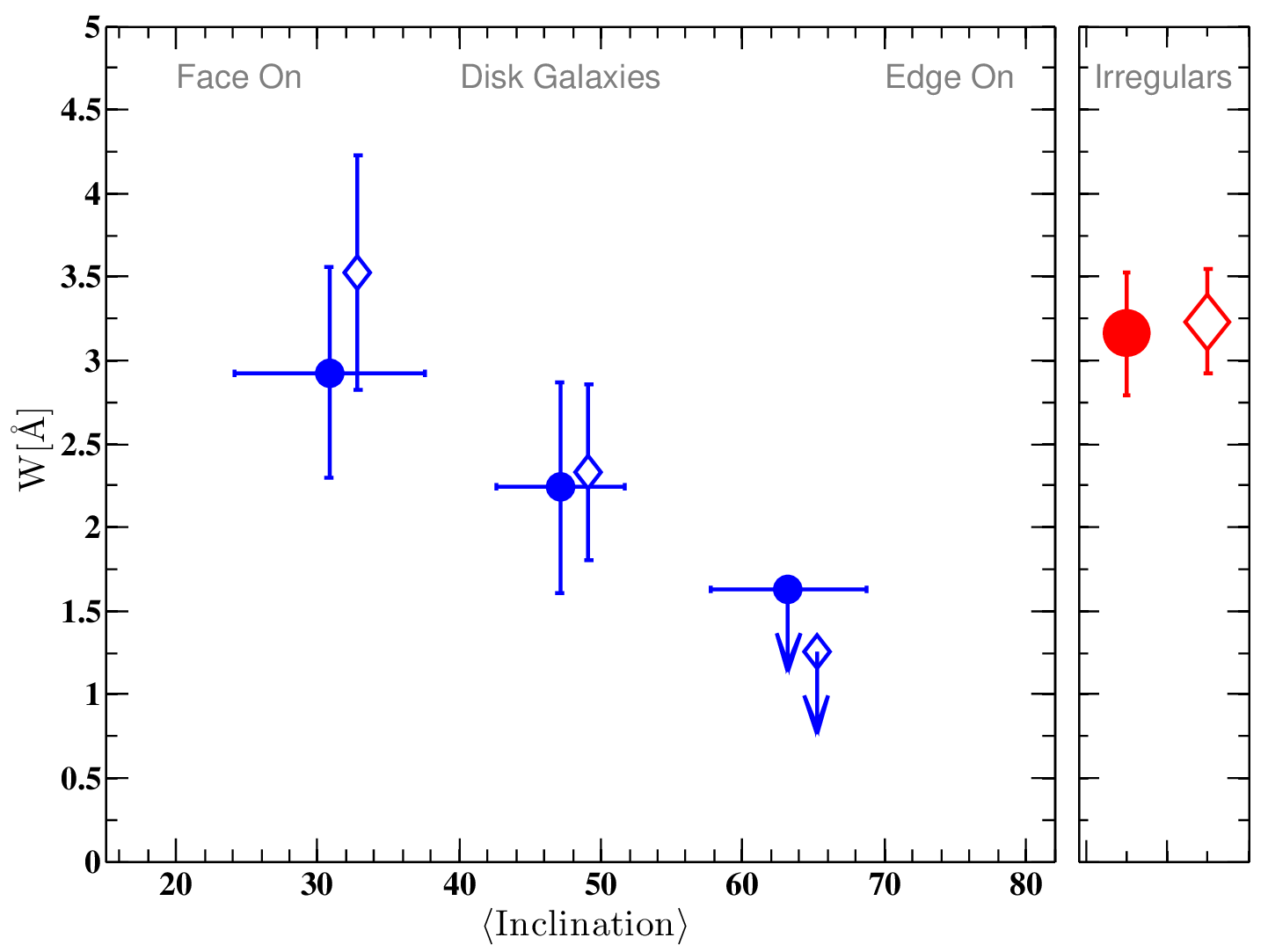}        
    \includegraphics[height=7.cm,width=8.5cm]{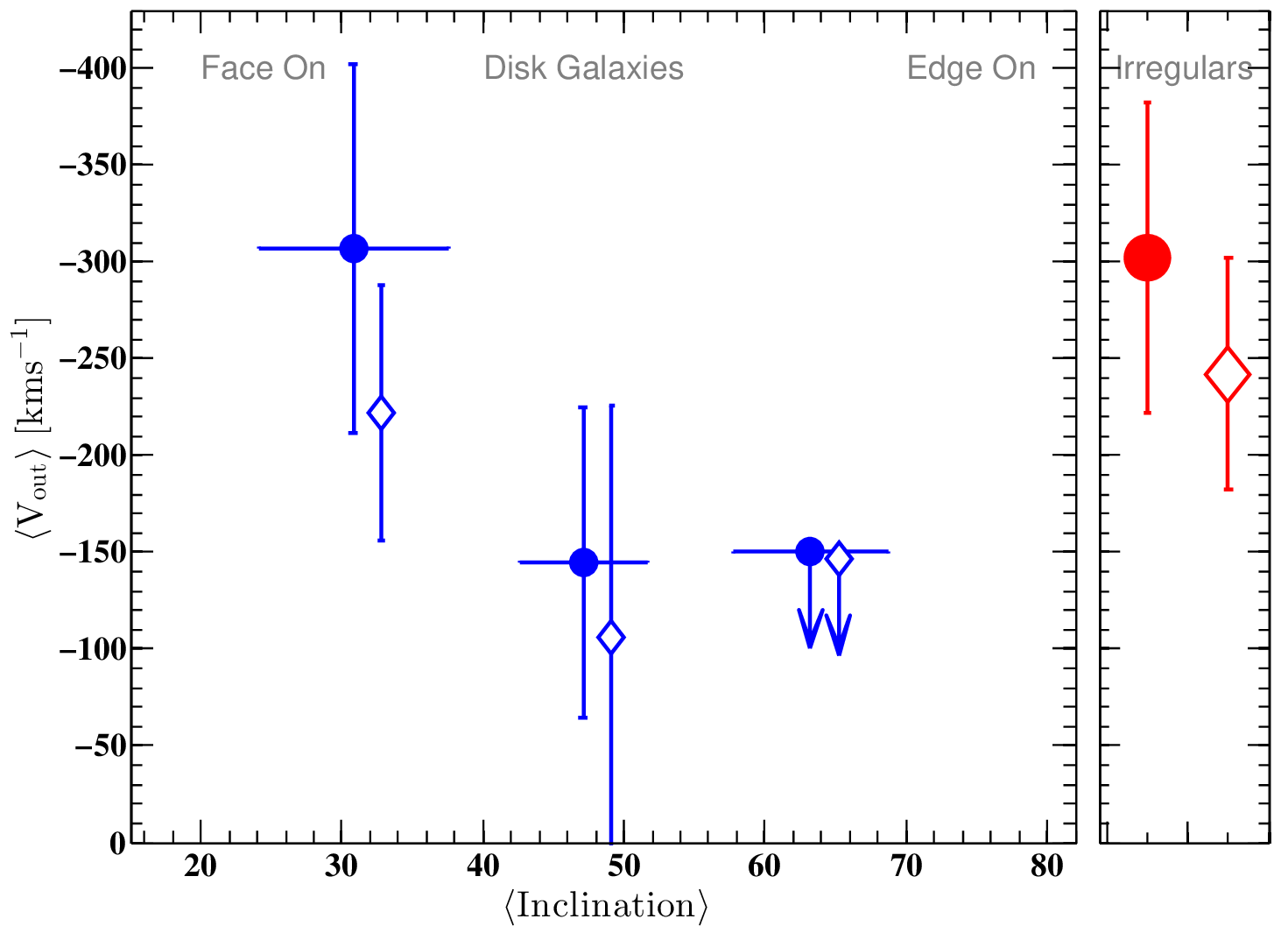}        
    \caption{Variation of outflow equivalent width (left panel) and outflow velocity (right panel) as a function of the apparent inclination of the disk galaxies. Open diamonds are boxcar measurements and the filled circles are decomposition measurements. Error bars on the x-axis are standard deviation in the inclination bin. The face-on galaxies clearly exhibit higher outflow equivalent widths and outflow velocities as compared to the edge-on galaxies. The red points are measurements for galaxies with irregular morphologies. The irregular galaxies exhibit as strong outflows as the face-on galaxies. The blue diamonds are offset in inclination angle for clarity.}
\label{ch5:fig12}
\end{figure*}

 Figure \ref{ch5:fig12} shows the variation of outflow equivalent width (left panel) with inclination and outflow velocity (right panel) with inclination. The error-bars in inclination angles are the standard deviation of the inclination angles within that bin. The face-on systems show the strongest outflow absorption equivalent width and this decreases with increasing inclination. The edge-on systems show the weakest outflow equivalent width and only a 1$\sigma$ upper limit was measured in both the decomposition (filled circles) and the boxcar (open diamonds) methods. The face-on systems also exhibit the highest outflow velocities and the observed outflow velocities decrease as we probe galaxies with higher inclinations. 

The galaxies with irregular morphologies, classified as ZEST type 3 have similar outflow equivalent widths as the face-on disk galaxies and similar outflow velocities ($\sim$ $-200 \; \rm{to} \;-300 \; \rm{kms^{-1}} $, red points figure \ref{ch5:fig12}). Since no inclination can be assigned to the galaxies with irregular morphologies, their outflow estimates are essentially averaged over all possible orientations. However, it should be noted that the mass distribution of these galaxies are not the same as that of the disk galaxies. A two sample KS test performed on the mass distribution of irregular and disk galaxies rules out the null hypothesis that they are drawn from the same mass distribution at 5\% significance level. To first order, they are shifted to lower mass by about 0.15 dex.

The trend of variation of outflow velocity with galaxy inclination is similar to the trends observed for low redshift samples.   \cite{chen_outflow2010} studied a sample of $\sim$ 150,000 SDSS galaxies and created stacks of galaxy spectra as a function of apparent disk inclination and studied the outflow properties of the Na D absorption doublet. They found a strong increase in outflow velocity as the galaxies became more and more face-on which is consistent with the picture of galactic winds escaping along the disk rotation axis. Further \cite{Heckman2000} observed Na D absorption in 18 local starbursts and found that a higher fraction of the face-on galaxies exhibit outflows as compared to the edge-on systems. These low redshift studies were feasible because of relatively bright targets and high resolution spectroscopy.

Recently, \cite{Wiener2009} examined 118 galaxies at $z \sim 1.4$ with HST I-band imaging and did not find a correlation between apparent inclination and wind strength or outflow velocity. They inferred that this might be due to uncertainties in estimating axis ratios of irregular galaxies imaged in the rest-frame U-band. \cite{Kornei2012} using a sample of  72 $z \sim 1$ galaxies, divided their sample at $i = 45^{\circ}$ and found that the co-added spectra of the face-on systems exhibit higher outflow velocity as compared to the more edge-on systems. 

Our results for inclination dependence of outflowing gas in the ``down the barrel'' spectra of disk galaxies are entirely consistent with the azimuthal dependence (around edge-on galaxies) of the distributed strong Mg II absorbers, seen against the spectra of background galaxies in \cite{Bordoloi2011a}. 

The new results tie directly to the picture that the azimuthal asymmetry of strong Mg II absorption line systems around galaxies, observed in the spectra of background galaxies or quasars, are primarily due to cool Mg II gas entrained in star formation driven bipolar winds, especially at low impact parameters (\citealt{Bordoloi2011a,Bordoloi2012a}).

\section{Minimum mass outflow rate}

The amount of gas ejected in galactic outflows is of great interest as it allows us to understand its importance to galaxy evolution (see e.g. \citealt{Lilly2013}). This is however not trivial to estimate. The Mg II lines observed are optically thick, and would be approaching the flat part of the curve of growth. Hence their column densities can not be inferred from their equivalent widths. We will, therefore estimate their apparent optical depths and hence apparent column densities which will give a conservative lower limit of the column density of MgII gas. The apparent optical depth of the line is given as

\begin{equation}
\tau_{a}(\lambda) \;=\; \ln \left[ \frac{I_{c}(\lambda)}{I(\lambda)} \right].
\end{equation}

The apparent column density per velocity element, $N_{a}(v)$ is given as
\begin{equation}
\log N_{a}(v) \;=\; \log \tau_{a}(v) - \log f \lambda_{0} + 14.576,
\end{equation}

where $\log f \lambda_{0} \;=\; 2.933$ for the Mg II 2803 line \citep{Morton1991}. We compute $\tau_{a}(v)$, between 0 and -768 $\rm{kms^{-1}}$ of the Mg II 2803 line. 

The integrated apparent column density $N_{a}$ which we take as the lower limit of the true column density of the outflowing gas is obtained as
\begin{equation}
 N_{a} \;=\; \int_{\rm{0\; kms^{-1}}}^{\rm{-768 \; kms^{-1}}} N_{a}(v) dv.
\end{equation}

We apply this procedure to estimate the column density of outflowing gas for the full sample of 486 galaxies and find $\log N(MgII) > 13.63$ and for the sample with highest $\Sigma_{SFR}$ we find $\log N(MgII) > 13.87$.

To estimate a conservative lower limit to $N_{H}$ in the outflow, we assume no ionisation correction (N(Mg) = N(MgII)), and take the minimum column density for Mg II with no correction for saturation. We assume solar abundance of Mg, $\log Mg/H \;=\; -4.42$ and assume a factor of -1.3 dex of Mg depletion onto dust \citep{Jenkins2009}. This gives a column density of  $\log N_{H} \geq 19.34$ for the whole sample and $\log N_{H} \geq 19.6$ for the highest $\Sigma_{SFR}$ bin. We stress that this estimate of column density of outflowing gas is a very conservative lower limit as we are neglecting ionization correction and underestimating any effects of saturation. Moreover, any redshifted emission from Mg II 2796 line will also reduce any blueshifted absorption from the Mg II 2803 line, further reducing the computed column densities here. 

The mass outflow rate for a wind with opening angle $\Omega_{w}$ is given as 
\begin{equation}
\rm{\dot{M}_{out} \;=\; \Omega_{w} \;C_{f} \;C_{\Omega}\; \mu m_{p}\; N_{H} \;R \;v,}
\end{equation}

where $C_{\Omega}$ is the angular covering fraction and $C_{f}$ is the clumpiness covering fraction,$\mu m_{p}$  is the mean atomic weight, $R$ is the minimum radius of the outflowing gas and $v$ is the mean velocity of the outflowing gas. Since the composite spectra of galaxies are viewed from all angles and the composites integrate over all small scale line of sight clumpiness as well as the partially and fully covered lines. Hence we set $C_{\Omega}C_{f}\; =\; 1$ and $\Omega_{w} \;=\; 4 \pi$ (thin shell approximation). Setting $\mu = 1.4$ we rewrite the mass outflow rate as (see \citealt{Wiener2009})

\begin{equation}
\rm{\dot{M}_{out} \;\simeq \; 22 M_{\odot} yr^{-1}  \frac{N_{H}}{10^{20} cm^{-2}}  \frac{R}{5 kpc}  \frac{v}{300 kms^{-1}}  }.
\end{equation}

We do not have any constraints on the spatial extent of the outflowing gas. However, it must be at least of the order of the size of the galaxies, since the covering factor is large. Hence we use the median half light radius of our galaxy sample as the minimum radial extent of the winds ($R  \sim $4.1 kpc).

Putting in all the values we find that for the whole sample the mass outflow rate is $\rm{\dot{M}_{out} \gtrsim 4.5 M_{\odot} yr^{-1}}$ and for the highest $\Sigma_{SFR}$ sample it is $\rm{\dot{M}_{out} \gtrsim 7 M_{\odot} yr^{-1}}$. These values are consistent with previous works \citep{Wiener2009,Rubin2010,Martin2005,Newman2012} and we find that $\rm{\dot{M}_{out}}$ is of the order of the star formation rate of the sample. Assuming that mass outflow rates are directly proportional to the star formation rate of the galaxies ($\rm{\dot{M}_{out}  \; = \; \eta SFR}$), we can put lower limits on the mass loading factor $\rm{\eta}$. We find that for the whole sample the average mass loading factor is $\rm{\eta} \gtrsim 0.24$. Again these values are conservative lower limits and better measurements on $R$ and $N_{H}$ are needed to constrain the mass outflow rates further.

\section{Conclusions}

In this work, we have analyzed the co-added spectra of ($\rm{R \sim 600}$) 486 zCOSMOS galaxies at $1 < z <1.5$ to study the properties of galactic outflows using Mg II absorption lines as tracers. We divided our sample in terms of the rest frame color, mass and star formation rates of the galaxies and studied the properties of outflow equivalent width and outflow velocity. Finally, we examined how outflow kinematics depend on the apparent inclination of disk galaxies and compared the findings to that of galaxies with irregular morphologies. The main findings of this study are as follows 

\begin{enumerate}
\item We find that the whole sample of 486 galaxies exhibits blueshifted absorption with a mean outflow velocity of $\rm{ v_{flow}  \; =\; -343 \pm 49 \; kms^{-1}}$ and $\rm{ v_{out}  \; =\; -319 \pm 95 \; kms^{-1}}$ measured with the decomposition and the boxcar method respectively. The total outflow equivalent width for the stacked spectrum is measured to be $W_{flow}\; =\; 2.55 \pm 0.24$ {\AA} and $W_{out}\; =\; 2.66\pm 0.25$ {\AA} respectively.

\item We find that the blue galaxies are associated with much stronger outflowing component as compared to the red galaxies in terms of their rest frame equivalent widths. At similar mean stellar masses  the blue galaxies exhibit almost four times higher outflow equivalent width as compared to the red galaxies.
 
\item Amongst the blue galaxies, outflow equivalent width also increases with increasing stellar mass. The uncertainties in outflow velocity estimates do not allow us to firmly establish an increase in outflow velocity with stellar mass of the host galaxies. Our findings also emphasize the need to consider the color of the galaxy in examining outflow trends with stellar mass or luminosity (as massive red galaxies show little outflow).

\item We find that galaxies with higher star formation rates exhibit higher outflow equivalent widths. There may be a correlation between outflow velocity with star formation rate, but due to large uncertainties in outflow velocity estimates this is of weak statistical significance.

\item Galaxies with high $\rm{\Sigma_{SFR}}$ exhibit strong outflow equivalent width as well as mean outflow velocity. We detected no outflow in the sample with $\rm{\Sigma_{SFR} \;  \leq \; 0.1 M_{\odot} yr^{-1} kpc^{-2}}$ consistent with the canonical $\Sigma_{SFR}$ threshold found at low redshifts.

\item Amongst the disk galaxies, the galaxies that are seen close to face on (inclination $i \leq 40^{\circ}$) are found to have $\sim$2.5 times higher outflow equivalent width as compared to the galaxies that are more close to being edge-on (inclination $i \geq 55^{\circ}$). In terms of outflow velocities, the face-on systems are also found to have higher outflow velocity ($\rm{\sim \; -200\; kms^{-1}}$ to $\rm{-300\; kms^{-1}}$) as compared to the edge-on systems, where only an upper limit of $\rm{\leq -150\; kms^{-1}}$ was detected. This dependence on  inclination suggests that for the disk galaxies, the galactic outflow is bipolar in nature and is primarily perpendicular to the disk of the galaxy.

\item For galaxies with irregular morphology, we find a relatively high outflow equivalent width and outflow velocity comparable to that of the face-on galaxies.

\item We also present lower limits on mass outflow rates ($\rm{\dot{M}_{out}}$) which for the whole sample is $\rm{\dot{M}_{out} \gtrsim 4.5 M_{\odot} yr^{-1}}$ and for the highest $\Sigma_{SFR}$ sample it is $\rm{\dot{M}_{out} \gtrsim 7 M_{\odot} yr^{-1}}$. Assuming $\rm{\dot{M}_{out} \propto SFR}$ we find the average mass loading factor $\rm{\eta}$ for the whole sample is $\rm{\eta} \gtrsim 0.24$.

\end{enumerate}

\section*{Acknowledgement}
RB would like to thank Jason Tumlinson for stimulating discussions and for carefully reading through the draft and giving insightful comments. This work has been partly supported by the Swiss National Science Foundation and is based on observations undertaken at the European Southern Observatory (ESO) Very Large Telescope (VLT) under Large Program 175.A-0839.

\begin{deluxetable*}{lcccccccc}
\tablewidth{0pt}
\tabletypesize{\footnotesize}
\tablecaption{EW measurements of the cladded spectra}
\tablehead{
\colhead{Sample}  & 
\colhead{$W_{flow}$\tablenotemark{a}} &
 \colhead{$W_{out}$\tablenotemark{b}} &
 \colhead{ $W_{all}$\tablenotemark{c}} &
 \colhead{ $\langle V_{flow} \rangle$\tablenotemark{d}} &
 \colhead{ $\langle V_{out} \rangle$\tablenotemark{e}} & 
 \colhead{$\langle V_{all} \rangle$\tablenotemark{f}} &
 \colhead{ No\tablenotemark{g}} &
 }
\startdata
All & $2.55 \pm 0.24$ & $2.66 \pm 0.25$ & $4.70 \pm 0.22$ & $-343 \pm 49 $ & $-319 \pm 95$ & $-179 \pm 51$ & 486 \\
Blue $\log{M_{*}} \geq 10.45 $ & $3.88 \pm 0.44$ & $3.77 \pm 0.45$ & $5.30 \pm 0.48$ & $-325 \pm 53$ & $-278 \pm 95$ & $-198 \pm 62$ & 126 \\
Blue $10 < \log{M_{*}} < 10.45 $ & $2.35 \pm 0.47$ & $2.14 \pm 0.49$& $4.80 \pm 0.34$ & $-330 \pm 61$ & $-290 \pm 120$ & $-130 \pm 45$ & 236 \\
Blue $\log{M_{*}} \leq 10 $ & $2.40 \pm 0.49$ & $2.82 \pm 0.5$ & $3.46 \pm 0.43$ & $-226 \pm 60$& $-110 \pm 80$ & $-87 \pm 50$ & 98 \\
Red & $< 1.8$ & $1.12 \pm .9$ & $ 6.12 \pm 0.86$ & < -106 & $ < -76$ & $60 \pm 80$ & 70 \\
$\log{SFR} \geq 1.5721 $ & $3.74\pm 0.50$ & $3.44\pm 0.49$& $6.01 \pm 0.45$ & $-299 \pm 95 $ & $-271 \pm 81$ & $-169 \pm 56$ & 119 \\
$1.0396 < \log{SFR} <1.5721 $ & $2.46\pm 0.30$& $2.94\pm 0.35$ & $5.47 \pm 0.29$ & $-263 \pm 100$ & $-230 \pm 99$ & $-125 \pm 86$ & 237 \\
$\log{SFR} \leq 1.0396 $ &$1.97 \pm 0.70$ & $2.09\pm 0.72$ & $4.42 \pm 0.74$ & $-152 \pm 105$ & $-179 \pm 118$ & $-80 \pm 50$ & 119 \\
$0 \leq \rm{inc} \leq 40$ & $2.92 \pm 0.63$ & $3.52 \pm 0.7$ & $4.32 \pm 0.47$ &$-307 \pm 95$ & $-222 \pm 66$ & $-180 \pm 37$ & 77 \\
$40 < \rm{inc} \leq 55$ & $2.24 \pm 0.62$ & $2.33 \pm 0.53$ & $4.92 \pm 0.42 $& $-145 \pm 80$ & $-106 \pm 120$ & $-50 \pm 40$ & 87 \\
$55 < \rm{inc} \leq 90$ & $<1.63$ & $<1.26$ & $4.77 \pm 0.50$ & $< -150$ & $<-147$ & $107 \pm 146$ & 54 \\
Irregulars & $3.16 \pm 0.37$ & $3.23 \pm 0.31$ & $5.18 \pm 35$ & $-302 \pm 80$ & $-242 \pm 60$ & $-150 \pm 35$ & 174 \\
$\log{\Sigma_{SFR}} \leq -1.04$ & $< 0.94 $ & $< 0.87$ & $4.69 \pm 0.73$ & --- & --- & $118 \pm 57$ & 103 \\
$-1.04 < \log{\Sigma_{SFR}} < -0.25$& $3.11 \pm 0.78$ & $3.07 \pm 0.47$ & $4.78 \pm 0.40$ & $-223 \pm 95$ & $-250 \pm 103$ & $-111 \pm 60$ & 207 \\
$\log{\Sigma_{SFR}} \geq -0.25 $& $3.97 \pm 0.58$ & $3.83 \pm 0.52$ & $5.64 \pm 0.47$ & $-257 \pm 50$ & $-254 \pm 90$& $-172 \pm 55$ & 103 \\
\enddata
\tablenotetext{a}{Outflow equivalent width from the decomposition method, in {\AA}}
\tablenotetext{b}{Outflow equivalent width from boxcar method, in {\AA}}
\tablenotetext{c}{Total equivalent width, in {\AA}}
\tablenotetext{d}{Mean outflow velocity from decomposition method, in $\rm{kms^{-1}}$}
\tablenotetext{e}{Mean outflow velocity from boxcar method, in $\rm{kms^{-1}}$}
\tablenotetext{f}{Mean absorption velocity, in $\rm{kms^{-1}}$}
\tablenotetext{g}{Number of galaxies}
\label{table:sample_table1}
\end{deluxetable*}

\renewcommand\bibsection{}
\bibliographystyle{thesis_bibtex}

\bibliography{mybibliography}

\end{document}